\documentclass{svjour3}                    
\usepackage{graphicx}
\usepackage{lineno}
\usepackage{setspace}
\usepackage{hyperref}
\hypersetup{
	colorlinks,
	citecolor=blue,
	filecolor=blue,
	linkcolor=blue,
	urlcolor=blue
}
\usepackage{amsfonts}
\usepackage{float}
\usepackage{xcolor}
\usepackage{cite}
\RequirePackage{amsmath}
\usepackage{multirow}
\usepackage{latexsym}

\journalname{Journal}
\begin{document}

\title{Mathematical Modeling of Blood Flow for a Diseased Model with Therapeutic Nanoparticles}

\titlerunning{Mathematical Modeling of blood flow for a Diseased Model} 

\author{Surabhi Rathore$^{1,\dagger}$ \and Dasari Srikanth$^{1,*}$}  

\authorrunning{Surabhi Rathore \and Dasari Srikanth} 

\institute{
    Corresponding Author${^*}$\\
	    \email{sri\_dasari1977@yahoo.co.in} \\
    \\
        ${^1}$  
        Department of Applied Mathematics, Defence Institute of Advanced Technology (Deemed to be University), Girinagar, Pune, Maharashtra, 411025, India\\
    \\
        Present Address${^{\dagger}}$\\
        mathLab, Mathematical Area, SISSA - International School of Advanced Studies, via Bonomea 265, Trieste, 34136, Italy\\         %
}
\date{Received: date / Accepted: date}

\maketitle

\begin{abstract}

The use of nanoparticles for targeted drug delivery, especially in diseased arteries, is a novel procedure. We are incorporating nanoparticles into blood vessels using a catheter, which could potentially deliver drugs precisely to affected areas, reducing side effects and increasing treatment efficiency. Considering non-Newtonian fluid modeling because blood is a complex fluid with non-linear behavior. In this paper, we are using mathematical modeling to understand blood flow dynamics, temperature, and concentration dispersion, which can provide valuable insights into the behavior of therapeutic nanoparticles in the bloodstream. The perturbation method is used to solve the complex mathematical model with permeable flow boundary conditions. We are investigating flow field characteristics including wall shear stress, pressure, and impedance to understand how nanoparticles disperse and interact with different physiological aspects. In conclusion, the proposed study focuses on the use of nanotechnology and mathematical modeling to understand the effects of therapeutic nanoparticles in diseased arteries, which is an important and valuable contribution to the medical field.



\end{abstract}
\section{Introduction}
\label{intro}
Nanotechnology greatly benefits the medical field, as it finds many applications in the diagnosis, treatment, and prevention of cardiovascular diseases (CVDs). Heart attacks and strokes, prevalent global health issues, often arise from physical abnormalities or malfunctions in the heart and blood vessels \cite{smith2023, ou2021}. In efforts to deliver nanomedicine precisely to affected regions, practitioners utilize cardiac catheterization, an invasive surgical technique~\cite{grossman1986}, which is employed to develop therapeutic approaches for direct drug administration to the affected site~\cite{mccarthy2010}.
CVDs are medical conditions caused by blood vessel and heart dysfunction, and include many physical conditions, such as rheumatic heart disease, ischemic heart disease, and cerebrovascular disease, as described in this report ~\cite{vos2020}. They are associated with high rates of mortality and morbidity, a significant health concern for modern society, and are accountable for $17.9$ million deaths each year, which is about $32\%$ of global deaths as stated in World Health Organization (WHO)~\cite{world2019}. In some medical conditions, the muscles of the blood vessels degenerate due to the accumulation of unsaturated fats, which is known as \textit{vasoconstriction}. When this situation occurs, blood flow to certain tissues in the human body is restricted. \textit{Vasodilation}, on the other hand, is the expansion of blood vessels resulting from the relaxation of the muscle walls of the blood vessel. This physiological phenomenon results in increased blood flow to parts of the body that lack oxygen or nutrients. According to these studies \cite{kellogg2006,mohrman1978,charkoudian2010}, dilation lowers blood pressure by reducing systemic resistance thereby increasing blood flow to affected blood vessels.

Numerous studies have been conducted to mathematically understand the effect of blood vessel abnormality on the flow dynamics within the lumen under different physiological conditions. Blood is a body fluid composed of blood cells, so several researchers modeled it as a non-Newtonian fluid by considering the movements of blood cells~\cite{fung1993}. Some theoretical non-Newtonian fluids replicating the behavior of blood are polar fluids represented by the couple-stress fluid~\cite{stokes2012} and the micropolar fluid~\cite{eringen1964}.
In these reports ~\cite{srinivasacharya2008,surabhi2021}, the blood flow characteristics for couple-stress fluid flow through constricted domains with different constraints are described. With the specified conditions, the wall shear stress (WSS) and impedance values for the non-Newtonian model were higher than those for the Newtonian fluid. Micropolar fluids are fluids with micro-structure embedded in them, was introduced by Eringen~\cite{eringen1964simple}. This model has a number of interesting advantages, beginning with the fact that it is a well-studied and extended generalization of the classical model. Some researchers \cite{ellahi2014,rathore2021,mekheimer2015suspension} have investigated blood flow dynamics within cylindrical geometry with mild stenosis assumptions, which was represented using a simple microfluid. These quantitative examinations using analytical and numerical approaches reveal reductions in flow velocities and flow rates. Whereas, when an artery narrows, the WSS rises, and in diverging scenarios, the pattern reverses. 

Nanofluid is a compound of nanoparticles (NPs) suspended in a base fluid that imposes the changes in their thermo-physical properties, which are thermal conductivity, convective heat transfer coefficients, viscosity, and thermal diffusivity~\cite{farokhzad2009, petros2010}. NPs have a significant potential in compounding blood physiology throughout storage, transport, and dispersion, according to~\cite{wen2009review, hatami2014computer}. Clinical trials have shown a reduction in re-narrowing when DEB is used, but the efficacy of this therapy is highly dependent on the drug's release rate and its distribution within the arterial lumen~\cite{chen2009predicting}. The novelty of NPs is well appreciated as similar scale base fluid particles do not have these properties, as mentioned in ~\cite{buongiorno2006convective}. Smallness and thermal fluctuations are the fundamental causes of nanoparticle Brownian motion, which affect thermal properties (thermal heat capacitance, specific heat capacitance, viscosity, etc.) as shown in table~\cite{surabhi2018impact, surabhi2021, elnaqeeb2016cu1}. NPs have a high tendency to compound the blood flow dynamics during storage, transportation, and dispersion. Understanding the rheological properties of nanofluids is crucial in determining their suitability for applications involving the convective transport phenomenon. Surabhi et al.~\cite{surabhi2018impact}  mathematically studied flow dynamics of blood for an $\omega-$ shaped catheterized stenotic geometry using varying slip velocities in restricted locations. These studies~\cite{nisar2020, ellahi2014, ahmed2017, shah2020} hold significant importance for pathology and the pharmaceutical industry, as it delves into the transport phenomena associated with drug delivery, motivated by the motion of NPs.

Recent advancements in the theoretical modeling of nanofluids have sparked considerable interest in understanding their thermophysical characteristics.  Therefore, this paper demonstrates the influence of NP migration due to the convective heat transfer parameters of nanofluids. We are critically understanding the therapeutic NPs behavior with physiological characteristics considered within the diseased model. 
\section{Problem Statement} 
\label{sec:2}
\subsection{\textbf{Mathematical flow geometry} }
\label{sec:2.1}
\begin{figure}[htbp]
\centering
\includegraphics[width=0.75\textwidth,height=1.8in]{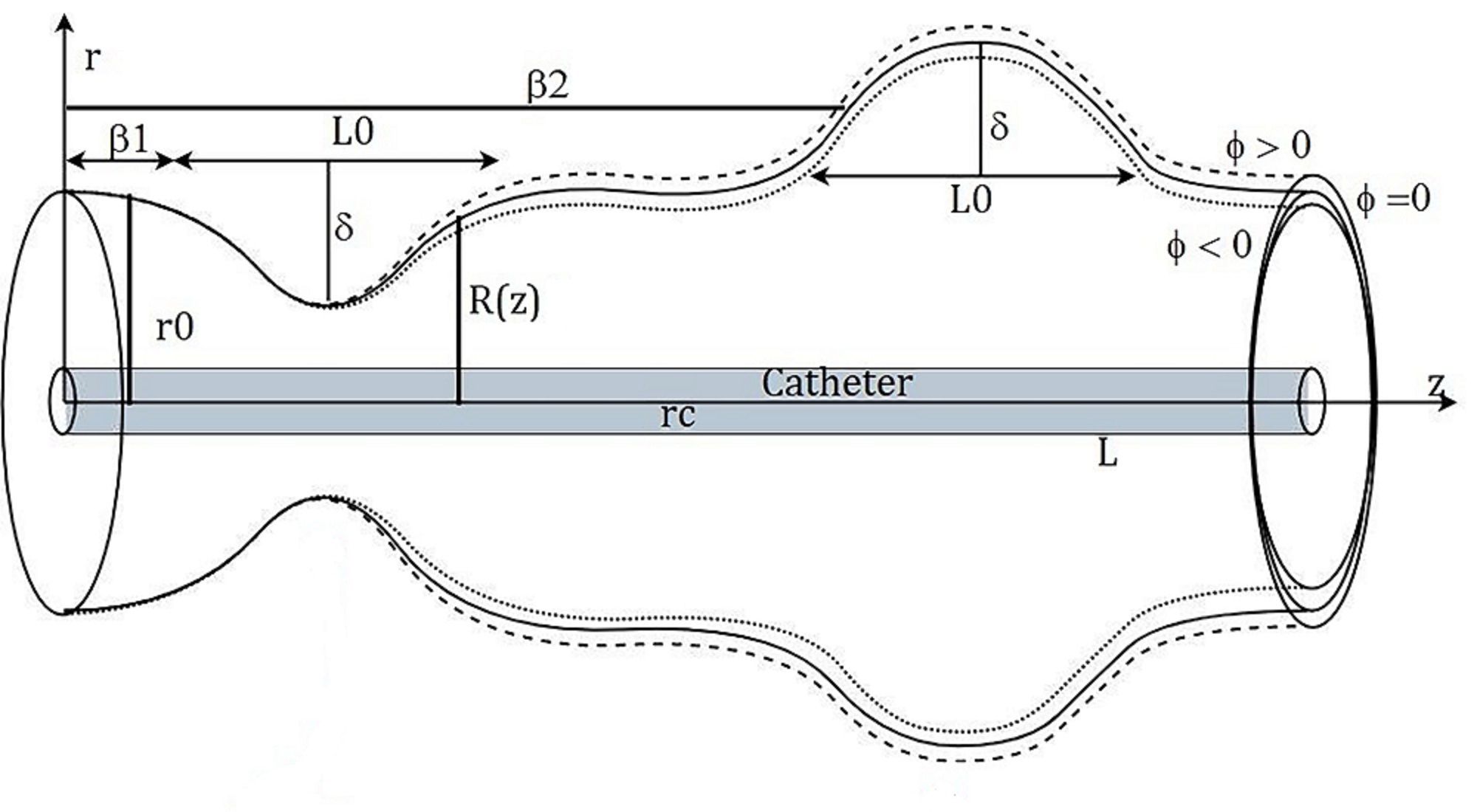}
\caption{Schematic representation of a vascular model.}
\label{fig1:geom}
\end{figure}

Considered axis-symmetric blood flow through a vascular diseased model with a catheter layered with therapeutic NPs, as depicted in Fig.~\ref{fig1:geom}. The blood flow is modeled as incompressible, laminar, and unsteady, and is represented by a non-Newtonian fluid. NPs are transferred into the bloodstream via a catheter with a layer of drug coated  on its outer surface. The resulting nanofluid flows within a region formed between the catheter $\left(r_c \right)$ and vascular wall with the varying radius $\left(R\left(z\right)\right)$. Here, $(r, \theta, z)$ are the cylindrical polar coordinates, where $r$, $\theta$, and $z$ respectively represent the radial, circumferential, and axial directions of the artery. Therefore, the considered flow domain is $\Omega \left(= [r_c, R(z)] \times [0, L] \right)$, and the mathematical expression for the same is taken as~\cite{pincombe1999effects} expressed as; 
\begin{equation}
 R(z) = \begin{cases} 
  \left( r_0 + \zeta z \right) \Big[ 1 - \frac{\delta}{2\,r_0} \Big( 1 + cos\frac{2 \pi}{l_i} \left( z - \beta_i - \frac{l_i}{2} \right) \Big)\Big], \\ & \hspace{-2in} \text{for \,\, $\beta_i \, \leq \, z \, \leq \, \beta_i + l_i$, \,\,  where \,\,$i\, = 1,\,2$;} \\
   \left(r_0 + \zeta z \right),  & \text{elsewhere.}
\end{cases}
\label{ch3_eq1} 
\end{equation}
Therein $l_i$ represents the length of the $i^{\mathrm{th}}$ abnormal segment, and $r_0$ represents the radius of the normal segment. $\displaystyle r_c$ is the catheter radius which is placed in the diseased vascular model and $\zeta (= tan\phi)$ represents tapering of the vessel. In this study, we consider the same lengths for the $i^{th}$ abnormal segment as $L_0$, which implies that $l_1 = l_2 = L_0$, and $\delta$ denotes the maximum height of both the vasoconstriction and vasodilation segments. $\beta_i$ are the locations of the $i^{\mathrm{th}}$ abnormal segment from origin respectively, which is expressed as $\displaystyle z_i = \left(\beta_i + {L_0}/{2}\right)$. The positive and negative values of $\delta$ in Fig.~\ref{fig1:geom}, correspond to the \textit{vasoconstriction} and \textit{vasodilation} segments of the considered  model, which is of length $L$.
\subsection{\textbf{Flow governing equations}}
\label{sec2.2}
The governing equations for an incompressible, laminar nanofluid,  with the pressure-gradient which is flowing through the domain as depicted in Fig.~\ref{fig1:geom}, are as follows:
\begin{equation}
\nabla \cdot \vec{V} = 0,
\label{ch2_eq8}
\end{equation}
\begin{equation}
\rho_{nf}\,\frac{D \vec{G}}{Dt} = - \nabla P + 2 \eta_{nf} \left( \nabla \times \vec{G} \right) + \left(\mu_{nf}+ \eta_{nf} \right) \nabla^{2}\,\left( \vec{V} \right) + F,
\label{ch2_eq9}
\end{equation}
\begin{equation}
\rho_{nf} \,J \,\frac{D \vec{G}}{Dt} = 2 \eta_{nf} \,\left( \nabla \times \vec{V}  \right) - 4 \eta_{nf} \left(\vec{G} \right) + 4 \Gamma \nabla^{2} \vec{G},
\label{ch2_eq10} 
\end{equation}
\begin{equation}
\frac{D\mathbb{T}}{Dt} = \frac{\kappa_{nf}}{(\rho c_{p})_{nf}} \nabla^{2} \mathbb{T} + \frac{\left(\rho c_{f} \right)_{nf}}{\left(\rho c_{p} \right)_{nf}} \left[ D_{B} \left(\nabla \mathbb{C} \cdot \nabla \mathbb{T} \right) +\frac{D_{T}}{\mathbb{T}_{1}} \left(\nabla \mathbb{T} \cdot \nabla T \right) \right], 
\label{ch2_eq11}
\end{equation}
\mbox{and}
\begin{equation}
\frac{D\mathbb{C}}{Dt}  = D_{B} \nabla^{2} \mathbb{C} + \frac{D_{T}}{\mathbb{T}_{1} }\nabla^{2} \mathbb{T}
\label{ch2_eq12}.
\end{equation}
Where, $\vec{V} = \left(u, 0, v \right)$ and $\vec{G} = \left(0, \omega, 0 \right)$ are bulk velocity and the angular momentum vectors respectively. Pressure-driven axisymmetric flow is considered in an artery of varying radius $R\left(z \right)$. $J$ represents the micro-inertia of blood cells and $\Gamma$ denotes the material constant. The Boussinesq approximation is used to approximate density variation in nanofluids, therefore $F$ which denotes the body force is governed by the buoyancy force expressed as 
\begin{equation}\label{ForceTerm}
\displaystyle F = g (\rho \gamma)_{nf} \left(\mathbb{T} - \mathbb{T}_{1} \right) + g(\rho \gamma)_{nf} \left(\mathbb{C}- \mathbb{C}_{1} \right).
\end{equation} 
$\rho_{nf}$ is the density and $P$ is the pressure of the nanofluid, respectively. $\mu_{nf}$ and $\eta_{nf}$ represent the dynamic and vortex viscosity of the considered nanofluid. $\mathbb{T}$ and $\mathbb{C}$ are temperature and concentration of bulk fluid. $D_{B}$ and $D_{T}$ represent the Brownian diffusion and thermophoretic diffusion coefficients respectively, which are expressed as, 
\begin{equation}
   D_B = \frac{K_{BO}\,\mathbb{T}}{3\,\pi\,\mu_f\,d_s}, \nonumber
\end{equation}
and 
\begin{equation}
    D_T =  0.26\,\left( \frac{\kappa_f}{2\,\kappa_f + \kappa_s}\right)\,\frac{\mu_f}{\rho_f}. \nonumber
\end{equation}
Here, $K_{BO}$ and $d_s$ represent the Boltzmann constant  and nanoparticle diameter, respectively. Thermal conductivity of nanofluid is $\kappa_{nf}$, specific heat capacitance of NPs is $\left(\rho c_{p} \right)_{nf}$ and heat capacitance of the nanofluid is $\left(\rho c_{f} \right)_{nf}$. The ratio of the thermal conductivity to the heat capacity of NPs in a fluid is $\displaystyle \alpha_{nf} = \frac{\kappa_{nf}}{\left( \rho c_{p} \right)_{nf}}$. Similarly, the ratio of heat capacities of the base fluid and that of NPs in the nanofluid is defined as $\displaystyle \tau_{nf} = \frac{\left(\rho c_{f} \right)_{nf}}{\left(\rho c_{p} \right)_{nf}}$.
Batchelor~\cite{batchelor1977effect} considered the bulk stress, which appeared because of Brownian motion in the solution, wherein the effective viscosity is considered as,
\begin{equation}
\frac{\mu_{nf}}{\mu_{f}} = \left(1+2.5\,\phi +6.25\,\phi^{2} \right).
\label{ch2_eq4}
\end{equation}
Thermophysical properties of nanofluids depend on volume fraction $\phi$, which follow the convective phenomena as discussed  in~\cite{buongiorno2006convective}.
 \begin{align}
\rho_{nf} &=  \rho_{s}\phi + \rho_{f}\left(1-\phi \right), \label{ch2_eq5}
\\
\left(\rho c_{p} \right)_{nf} &=  \left(\rho c_{p} \right)_{s}\phi + \left(\rho c_{p} \right)_{f}\left(1-\phi \right),\label{ch2_eq6}
\\
\left(\rho \gamma \right)_{nf} &=  \left(\rho \gamma \right)_{s} \phi + \left(\rho \gamma \right)_{f} \left(1-\phi \right). \label{ch2_eq7}
\end{align}   
Here, the subscripts $f$, $s$, and $nf$ respectively stand for base fluid,  NPs, and nanofluid.
\\
The non-dimensional parameters are introduced as follows;
\begin{align}
\begin{split}
r'=\frac{r}{r_0}, \hspace{0.2cm} z'=\frac{z}{L_1}, \hspace{0.2cm} P'=P\,\frac{r_{0}^{2}}{(u_0\,L_1\,\mu_{f})}, \hspace{0.2cm} t'=t\,\frac{u_0}{L_1}, \hspace{0.2cm} 
v'= \frac{v}{u_0},\\ \hspace{0.2cm} \Theta = \frac{\left( \mathbb{T} - \mathbb{T}_1 \right)}{\left( \mathbb{T}_0 - \mathbb{T}_1 \right)}, \hspace{0.2cm} \varsigma = \frac{\left( \mathbb{C} - \mathbb{C}_1 \right)}{\left( \mathbb{C}_0 - \mathbb{C}_1 \right)}, \hspace{0.2cm}
u'= \frac{u\,L_1}{\left(u_0\,R_{\epsilon} \right)}, \hspace{0.2cm} \omega' = \frac{ r_0\,\omega}{u_0}, \hspace{0.2cm} J' = \frac{J}{ r_{0}^{2}}.
\end{split}
\label{ch2_eq19}
\end{align}
Where $r_{0}$ and $u_{0}$ are the characteristic radius and velocity of the normal segment, respectively. $\displaystyle Re = \frac{\rho_{f} u_{0}r_{0}}{\mu_{f}}$ is the Reynolds number. This study considers the flow geometry to be far off from the heart, therefore, the flow is modeled as a low-Reynolds-number flow.

Introduction of the non-dimensional parameters into the governing Eqs. (\ref{ch2_eq8}) -- (\ref{ch2_eq12}), results in the following equations ( primes $(')$ are dropped) ,
\begin{equation}
\xi\,\Big({\partial_{r} u} + \frac{u}{r} \Big) + {\partial_z v} = 0,
\label{ch2_eq20}   
\end{equation}
\begin{multline}
\displaystyle \frac{\rho_{nf}}{\rho_{f}}\,Re\,\xi\delta^{3}\,\Big({\partial_{t} u} + \xi u {\partial_r u} + v {\partial_z u} \Big) = - {\partial_r P} -2\,\frac{N}{(1-N)}\,C_{\eta}\,\delta^{2}\,{\partial_z \omega} \\ + \left(C_{\mu} + \frac{N}{(1-N)} C_{\eta} \right)\,\delta^{2}\,\xi \,\left({\partial_{rr} u} + \frac{1}{r} {\partial_r u} - {u}/{r^{2}} +\delta^{2}\,{\partial_{zz} u} \right),
\label{ch2_eq21}  
\end{multline}
\begin{multline}
\displaystyle \frac{\rho_{nf}}{\rho_{f}}\,Re\,\xi\,\delta\, \Big( {\partial_t v} +  \xi u {\partial_r v} + v {\partial_z v}\Big) =  - {\partial_z P} + 2\,\frac{N}{(1-N)}\,C_{\eta}\,\left({\partial_r \omega} + \frac{\omega}{r} \right) \\ + \left( C_{\mu}+\frac{N}{(1-N)}\,C_{\eta} \right)\,\left( {\partial_{rr} v} + \frac{1}{r}\,{\partial_{r} v} \right) + \frac{(\rho \gamma)_{nf}}{(\rho \gamma)_{f}}G{r}\Theta + \frac{(\rho \gamma)_{nf}}{(\rho \gamma)_{f}}B{r}\varsigma,
\label{ch2_eq22}    
\end{multline}
\begin{multline}
\displaystyle \frac{\rho_{nf}}{\rho_{f}}\,Re\, J \,\delta \Big( {\partial_t \omega} + \xi\, u\, {\partial_r \omega} + v {\partial_z \omega} \Big) = 2C_{\eta}\,\frac{N}{(1-N)}\,\left(\delta \, {\partial_z v} - {\partial_r v} \right) - 4\frac{N}{(1-N)}\,C_{\eta}\,\omega \\ + 2\,\frac{(2-N)}{M^{2}}\,\left( {\partial_{rr} \omega} + \delta^{2}{\partial_{zz} \omega} + \frac{1}{r}\,{\partial_r \omega}- \frac{\omega}{r^{2}} \right),
\label{ch2_eq23}    
\end{multline}
\begin{multline}
\displaystyle Re \,Pr \,\delta \,\Big({\partial_t \Theta} + \xi u {\partial_r \Theta} + v {\partial_z \Theta} \Big) = \left({\partial_{rr} \Theta} + \frac{1}{r}\,{\partial_{r} \Theta} + \delta^{2}\,{\partial_{zz} \Theta} \right) \\ + N_{b}\,\left({\partial_r \Theta}\,{\partial_r \varsigma} + \delta^{2} {\partial_z \Theta}\,{\partial_z \varsigma} \right) + N_{t}\,\left( \left({\partial_r \Theta}\right)^{2} + \delta^{2} \left({\partial_z \Theta} \right)^{2} \right), 
\label{ch2_eq24}   
\end{multline}
\mbox{and}
\begin{multline}
\displaystyle Re\,\delta \,Pr\,\Big( {\partial_t \varsigma} + \xi\, u \,{\partial_r \varsigma} + v \,{\partial_z \varsigma} \Big) = \left({\partial_{rr} \varsigma}+ \frac{1}{r}\,{\partial_r\varsigma} + \delta^{2}\, {\partial_{zz} \varsigma} \right)  \\ + \frac{N_{t}}{N_{b}}\,\left({\partial_{rr} \Theta} + \frac{1}{r}\,{\partial_r \Theta} + \delta^{2}\,{\partial_{zz} \Theta} \right). 
\label{ch2_eq25} 
\end{multline}
We consider dynamic viscosity and micro-rotation viscosity of the nanofluid of the form $\mu_{nf} = \mu_{f}C_{\mu}, \hspace{0.2cm} \mbox{and} \hspace{0.2cm} \eta_{nf} = \eta_{f}C_{\eta},$ where  $\mu_{f}$  and $\eta_{f}$ are the dynamic viscosity and micro-rotation viscosity of the micro-polar fluid, and $C_{\mu}$ and $C_{\eta}$ are the functions of the nanoparticle volume fraction. The micropolar parameter is denoted by $ \displaystyle  M^{2} = \frac{r_{0}^{2}\eta_{f}\left[2\mu_{f}+\eta_{f}\right]}{\Gamma\,\left[\mu_{f}+\eta_{f}\right]}$  and $\displaystyle N = \frac{\eta_{f}}{\left(\mu_{f} + \eta_{f}\right)}$ is the coupling number. 

Under the assumptions of mild stenosis, i.e., $\xi\,\left(= R_{\epsilon}/r_{0} \right) << 1 $ with the additional geometric condition $\delta(= r_{0}/L_{1}) \approx o(1)$, the resultant non-dimensional governing Eqs. (\ref{ch2_eq20}) - (\ref{ch2_eq25}) are as given below,
\begin{equation}
{\partial_z v} = 0,
\label{ch2_eq26}
\end{equation}
\begin{equation}
{\partial_r P} = 0,
\label{ch2_eq27}
\end{equation}
\begin{multline}
\displaystyle {\partial_z P} = \left( C_{\mu}+\frac{N}{(1-N)}C_{\eta} \right)\,\left( {\partial_{rr} v} +\frac{1}{r}\,{\partial_r v} \right) + 2\left(\frac{N}{(1-N)} \right) \,C_{\eta}\,\left( {\partial_r \omega} + \frac{\omega}{r} \right) \\ + \beta_{nf} G{r}\Theta + \beta_{nf} B{r}\,\varsigma,
\label{ch2_eq28}   
\end{multline}
\begin{equation}
\displaystyle \frac{(2-N)}{M^{2}}\,\left( {\partial_{rr} \omega} + \frac{1}{r}\,\frac{\partial_r \omega} - \frac{\omega}{r^{2}}  \right)  =  C_{\eta}\,\frac{N}{(1-N)}\,\left({\partial_r v}+ 2 \omega \right),
\label{ch2_eq29}    
\end{equation}
\begin{equation}
\displaystyle \left( {\partial_{rr} \Theta} + \frac{1}{r}\, {\partial_r \Theta}\right) + N_{b}\,\left({\partial_r \varsigma}{\partial_r \Theta} \right)+ N_{t} \left( {\partial_r \Theta} \right)^{2} = 0, 
\label{ch2_eq30}    
\end{equation}
\begin{equation}
\displaystyle \left( {\partial_{rr} \varsigma} + \frac{1}{r}\, {\partial_r \varsigma} \right) + \frac{N{t}}{N{b}} \left({\partial_{rr} \Theta}  + \frac{1}{r}\, {\partial_r \Theta} \right) = 0.
\label{ch2_eq31}    
\end{equation}
Here, $\beta_{nf} \Big(:= \frac{(\rho \gamma)_{nf}}{(\rho \gamma)_{f}}\Big)$ is the ratio of the thermal expansion coefficients of the nanofluid concerning the micropolar fluid, $\displaystyle G{r} = \frac{r_{0}^{2}g(\rho \gamma)_{f}(\mathbb{T}_{0} - \mathbb{T}_{1})}{u_{0}\mu_{f}}$ is the Grashof number and $\displaystyle B{r} = \frac{r_{0}^{2}g(\rho \gamma)_{f}(\mathbb{C}_{0} - \mathbb{C}_{1})}{u_{0}\mu_{f}} $ is corresponding to the solutary Grashof number. $\displaystyle N_{b} = \frac{(\rho c_{f})_{nf}D_{B}( \mathbb{C}_{0} - \mathbb{C}_{1})}{(\rho c_{p})_{nf}}$ is known as Brownian motion parameter and  $\displaystyle N_{t} = \frac{(\rho c_{f})_{nf}D_{T}(\mathbb{T}_{0} - \mathbb{T}_{1})}{\mathbb{T}_{1}(\rho c_{p})_{nf}}$ as the thermophoresis parameter.
\subsection{\textbf{Physiological Boundary conditions}}
\label{sec2.3}
Rheological properties of blood are very important, therefore the flow characteristics in the considered domain indeed depend on the boundary conditions. As the  catheter's outer surface, which is layered with NPs, is being  placed inside the lumen of the blood vessel, temperature, and concentration boundary conditions are given as;
\begin{align}
    \Theta = 1, \hspace{2mm} \varsigma = 1, \hspace{2mm} \mbox{at} \hspace{2mm} r = r_c,  \label{bc:temp}
    \\
     \Theta = 0, \hspace{2mm} \varsigma = 0, \hspace{2mm} \mbox{at} \hspace{2mm} r = R(z). \label{bc:conc}
\end{align}
Boundary conditions at the catheter wall are,
\begin{equation}
    v = u_0, \hspace{2mm} \omega = 0, \hspace{2mm} \mbox{at} \hspace{2mm} r = r_c. \label{bc:vel1}
\end{equation}
Brunn~\cite{brunn1975} understood the presence of slip velocity at the vessel wall for polar fluids, which are expressed as
\begin{align}
    v = u_{slip}, \hspace{2mm} \mbox{at normal segments of} \hspace{2mm} R(z),\label{bc:vel2}
    \\
    \displaystyle  {\partial_r v} = \frac{\chi}{\sqrt{D_a}} \Big( u_{slip} - u_{porous}\Big), \mbox{at \textit{vasoconstriction} segment of} \hspace{2mm} R(z), \label{bc:vel3} 
    \\
    \displaystyle {\partial_r v} = \frac{\chi}{\sqrt{D_a}} \Big( u_{slip} + u_{porous}\Big), \mbox{at \textit{vasodilation} segment of} \hspace{2mm} R(z), \label{bc:vel4} 
    \\
    \omega = -\frac{n_1}{2} {\partial_r v} , \hspace{2mm} \mbox{at} \hspace{2mm} r = R(z), \hspace{2mm} \mbox{at} \hspace{2mm} 0 \leq n_1 \leq 1 ,  \label{bc:vel5}
\end{align}
where $u_{slip}$ is slip velocity in the normal segments and $\displaystyle u_{porous} = -\frac{D_a}{\mu_f} ,{\partial_z P}$. The Darcy number is $D_a$, and $\chi$ is a constant quantity that depends on the porosity of the material structure near the boundary.
\\
\section{Method of solution}
\label{sec:3}
In 1999, J.H. He developed the homotopy perturbation method (HPM) to solve the nonlinear partial differential equations (PDEs)~\cite{he2000variational}, allowing the freedom to choose the linear operator and the initial approximation. By employing this flexible approach, nonlinear coupled PDEs can be transformed into a system of solvable PDEs with appropriate boundary conditions. One of the notable strengths of the HPM is its efficiency in achieving accurate solutions with a small number of iterations. The convergence of the method sets it apart from other perturbation methods, making it an effective computational tool. In this study, to compute the solution of the nonlinear governing Eqs. (\ref{ch2_eq26}) -- (\ref{ch2_eq31}), we adopted this method.
\subsection{\textbf{HPM Procedure}}
\label{sec3.1}
Considering the following non-linear differential equation with boundary condition;
\begin{align}
 D(U) = f(r), \hspace{2cm} r \in  \Omega \label{ch2_eq43}
\\
B(U, \partial_n U) = 0,  \hspace{2cm} r \in  \Gamma_h \label{ch2_eq44}
\end{align}
where $D$, $B$, and $f(r)$ are the non-linear differential operator, boundary operator, and given analytic function. $\Omega$ is the considered domain and  $\Gamma_h$ is the boundary of $\Omega$. $D$ can be divided into $L$ , a linear operator, and $N$ , a non-linear operator. Therefore, the differential Eq. (\ref{ch2_eq43}) is rewritten as:
\begin{equation}
L(U) + N(U) - f(r) = 0. \label{ch2_eq45}
\end{equation}
We construct a homotopy following the approach proposed by Liao \cite{liao2011homotopy} as, $W(r, p): \Omega \times [0, 1] \rightarrow \mathbf{R}^n$, which satisfies
\begin{equation}
H(W, p) = (1 - p)\left[L(W) - L(U_0)] + p[D(W) - f(r)\right] = 0, \label{ch2_eq46}
\end{equation}
where $p \in [0, 1]$ is the perturbation parameter and $U_0$ is the solution of the linear operator which satisfies the boundary condition and is also treated as the initial solution of Eq. (\ref{ch2_eq43}). Furthermore, from Eq. (\ref{ch2_eq46}), it is evident that;
\begin{equation}
   H(W, 0) = L(W) - L(U_0) = 0,  \label{ch2_eq47} 
\end{equation}
\begin{equation}
   H(W, 1) = D(W) - f(r) = 0.  \label{ch2_eq48} 
\end{equation}
$W(r, p)$ varies from $U_0(r)$ to $U(r)$ as the value of perturbation parameter $p$ changes from zero to one, which is called homotopy. Consider a power series solution for variable $W(r,p)$ in the increasing  power of $p$ given as;
\begin{equation}
W(r,p) = W_0 + p\,W_1 + p^2\,W_2 + p^3\,W_3 + \cdots.  \label{ch2_eq49}
\end{equation}
The semi-analytic solution of $U$ can be obtained by putting $p=1$, in the above equation
\begin{equation}
U(r) = \lim_{p\rightarrow 1}W = W_0 + W_1 + W_2 + W_3 + \cdots.  \label{ch2_eq50}
\end{equation}
This process is called HPM.
\subsection{\textbf{Convergence of HPM}}
\label{sec3.2}
The convergence of HPM for PDEs with nonlinear operators has been discussed in~\cite{biazar2009convergence}. Let us rewrite Eq. (\ref{ch2_eq46}) in the following form;
\begin{equation}
L(W) = L(U_0) + p\left[ f(r) - N(W) - L(U_0)\right] = 0, \label{conv1}
\end{equation}
We obtain the following by applying the inverse operator $L^{-1}$ to both sides of Eq. (\ref{conv1}),
\begin{equation}\label{conv2}
W = U_0 + p\left[L^{-1}f(r) - \left(L^{-1}N\right)W - U_0\right].
\end{equation}
The perturbation solution of the variable $U$ is as follows
\begin{equation}\label{conv3}
W = \sum_{i=0}^{\infty}\,p^{i}\,W_i,
\end{equation}
substituting Eq. (\ref{conv3}) into the right hand side of Eq. (\ref{conv2}), we obtain
\begin{equation}\label{conv4}
W = U_0 + p\left[L^{-1}f(r) - \left(L^{-1}N \right)\,\left( \sum_{i=0}^{\infty}\,p^{i}\,W_i,\right) - U_0\right].
\end{equation}
The solution is obtained by utilizing the condition, $p\rightarrow 1$ as
\begin{equation}\label{conv5}
U =  \lim_{p\rightarrow 1}W
= \left[L^{-1}f(r) - \left(L^{-1}N \right)\,\left( \sum_{i=0}^{\infty}\,p^{i}\,W_i,\right)\right].
\end{equation}
We state the following Theorem as available in literature related to convergence.
\begin{theorem} \textbf{Sufficient Condition for convergence}
\\
Suppose that $X$ and $Y$ are Banach spaces and $N : X \rightarrow {Y}$ is a constrictive nonlinear mapping, that is
\begin{equation}\label{conv6}
    \forall v, \,\,\,v^{*} \in X;  \,\,\, \| N(v) - N(v^{*}) \|\, \leq \alpha \,\| v  - v^{*} \|, \,\,\,\,\, 0 \, <\, \alpha \, < \,1.
\end{equation}
Then according to Banach’s fixed point theorem operator $N$ has a unique fixed point $x$, that is $N(x) = x.$
Assume that the sequence generated by the homotopy perturbation method can be written 
\begin{equation}\label{conv7}
V_n = N(V_{n-1}); \,\,\,\,\, V_{n-1} = \sum_{i=0}^{n-1}v_i; \,\,\,\, n = 1,2,3,\cdots,
\end{equation}
and suppose that $V_0 = v_0 \in B_r(v)$  where $B_r(v)= \{ v^{*} \in X;\,\, \| (v - v1) \| < r \} $ then we have

\begin{enumerate}
    \item $V_{n} \in B_r(v),$
    \item $\lim_{n\rightarrow \infty} V_n = V.$
\end{enumerate}
\end{theorem}
\subsection{\textbf{Implementation of HPM}}
\label{sec3.3}
For solving the nonlinear PDEs (\ref{ch2_eq26}) -- (\ref{ch2_eq31}), we adopt the HPM method, as discussed above. The chosen linear operator is given as:
\begin{equation}
L = \left( {\partial_{rr} }+ \frac{1}{r}\,{\partial_r } \right) . \label{eq:iplm1}
\end{equation}
We also consider the power series for the variable $S\subset$ $\left( \Theta, \varsigma, v, \omega \right)$ with the perturbation parameter $p \in (0,1)$,  satisfying the prescribed boundary conditions (\ref{bc:temp}) -- (\ref{bc:vel5}). 
\begin{align}
S_h(r,p) &= S_0 + p S_1 + p^2 S_2 + p^3 S_3 + \cdots,
\label{eq:iplm2}
\end{align}
\\
Considered power series satisfy Eq. (\ref{ch2_eq48}), and results in;
\begin{multline} \label{eq:iplm6}
\displaystyle H(p, \Theta_h) = (1-p)\,\left[ L(\Theta_h) - L(\Theta_{0}) \right] + p\,\left[ L(\Theta_h) + N_{b}\,\big( {\partial_r \Theta_h}\,{\partial_r \varsigma_h} \big) + N_{t}\,\big({\partial_r \Theta_h}\big)^{2} \right] , 
\end{multline}
\begin{multline}\label{eq:iplm7}
\displaystyle H(p, \varsigma_h) = (1-p)\,\left[ L(\varsigma_h) - L(\varsigma_{0}) \right] + p \Bigg[ L(\varsigma_h) + \frac{N_{t}}{N_{b}}\,\left( \frac{1}{r} {\partial_r }\,\left( r {\partial_r \Theta_h}\right) \right) \Bigg] , 
\end{multline} 
\begin{multline}
\displaystyle H(p, v_h)  = (1-p)\big[L(v_h) - L(v_{0})\big] + p\Bigg[ L(v_h) +\frac{2\,N\,C_{\eta}}{(1-N)\,C_{\mu}+NC_{\eta}}\Big( \frac{\partial \omega_h}{\partial r}+ \frac{\omega}{r}\Big) \\ + \frac{(1-N)}{(1-N)C_{\mu} + N\,C_{\eta}}
  \Big( \beta_{nf}\,G_{r}\Theta_h +  \beta_{nf}B_{r}\,\varsigma_h - {\partial_z P} \Big)\Bigg],  
\label{eq:iplm8}
\end{multline}
and,
\begin{multline}
\displaystyle H(p, \omega_h) = (1-p)\big[L(\omega_h) -L(\omega_{0})\big] + p\Big[ L(\omega_h)-\frac{C_{\eta}M^{2}}{(2-N)}\frac{N}{1-N}\Big( {\partial_r v_h} +2 \omega_h \Big) -\frac{\omega_h}{r^{2}} \Big]. 
\label{eq:iplm9}   
\end{multline}
The above equations are simplified as; 
\begin{multline}
\displaystyle H(p, \Theta_h) = \left( \frac{1}{r}\,{\partial_r } \left( r\,{\partial_r \Theta_h} \right) \right) - \left( \frac{1}{r}\,{\partial_r }\,\left( r {\partial_r \Theta_{0}} \right) \right) + p \,\left( \frac{1}{r}\,{\partial_r }\,\left( r\,{\partial_r \Theta_{0}} \right) \right) \\ + p \left[ N_{b} \left( {\partial_r \varsigma_h}\,{\partial_r \Theta_h} \right) + N_{t}\left({\partial_r \Theta_h} \right)^{2} \right],
\label{eq:iplm10}    
\end{multline}
\begin{multline}
\displaystyle H(p,\varsigma_h) = \left( \frac{1}{r}{\partial_r }\,\left( r{\partial_r \varsigma_h} \right) \right) - \left( \frac{1}{r}{\partial_r }\,\left( r {\partial_r \varsigma_{0}} \right) \right) + p\, \left( \frac{1}{r}{\partial_r }\,\left( r {\partial_r \varsigma_{0} }\right) \right) \\ + p \,\left[\frac{N_{t}}{N_{b}}\, \left( \frac{1}{r}{\partial_r}\,\left( r {\partial_r \Theta_h} \right) \right)  \right], 
\label{eq:iplm11}    
\end{multline}
\begin{multline}
\displaystyle H(p, v_h)  = \left( \frac{1}{r}{\partial_r} \left( r {\partial_r v_h} \right) \right) - \left( \frac{1}{r} {\partial_r } \left( r {\partial_r v_{0} } \right) \right) + p\,\left( \frac{1}{r}{\partial_r } \Big( r {\partial_r v_{0} } \Big) \right) \\+ p\Bigg[\frac{2\,N\,C_{\eta}}{(1-N)C_{\mu}+NC_{\eta}}\Big( {\partial_r \omega_h} + \frac{\omega_h}{r}\Big) + \frac{(1-N)}{(1-N)C_{\mu} + N C_{\eta}}\Big( \beta_{nf}\,G_{r}\Theta_h + \beta_{nf}B_{r}\varsigma_h - {\partial_ P}\Big)\Bigg], 
\label{eq:iplm12}
\end{multline}
and
\begin{multline}
\displaystyle H(p, \omega_h) = \left( \frac{1}{r} {\partial_r }\Big( r{\partial_r \omega_h } \Big) \right)- \left( \frac{1}{r}{\partial_r}\Big( r {\partial_r \omega_{0} } \Big) \right) + p\, \left( \frac{1}{r} {\partial_r } \Big( r {\partial_r \omega_{0}}\Big) \right) \\ + p\Bigg[ - \frac{C_{\eta}M^{2}}{(2-N)}\frac{N}{1-N}\Big( {\partial_r v_h} + 2\omega_h \Big)  -\frac{\omega_h}{r^{2}} \Bigg],
\label{eq:iplm13} 
\end{multline}
where $p$ is a perturbation parameter that has the range $0 \leq p \leq 1$.
\\
At $p=0$, we obtained the initial solutions for all the variables, considering the boundary conditions. Here are the initial solutions for the temperature and concentration profiles: 
\begin{align} 
    \Theta_0 &= \frac{\Big({\log(R(z)-\log(r)}\Big)}{\Big({\log(R(z)-\log(r_c)}\Big)},\hspace{10mm} \mbox{and} \hspace{10mm}
    \sigma_0 &= \frac{\Big({\log(R(z)-\log(r)}\Big)}{\Big({\log(R(z)-\log(r_c)}\Big)}.\label{eq:iplm36} 
\end{align}
Initial solutions for axial velocity profiles for various segments are,
\begin{align}
  \displaystyle v_0 &= \frac{\Big[ u_0 \Big({\log(R(z)-\log(r)}\Big) + u_s \Big({\log(r)-\log(r_c)}\Big) \Big]}{\Big({\log(R(z)-\log(r_c)}\Big)}, \,\, \mbox{normal segments} \label{eq:iplm38} 
  \\
    v_0 &= \Big[ u_0 + R(z)\,\frac{\chi}{\sqrt{D_a}} \Big( u_{slip} - u_{porous}\Big) \Big({\log(r)-\log(r_c)}\Big) \Big], \,\, \mbox{ vasoconstriction} \label{eq:iplm39}
  \\
   v_0 &= \Big[ u_0 + R(z)\,\frac{\chi}{\sqrt{D_a}} \Big(u_{slip} + u_{porous}\Big) \Big({\log(r)-\log(r_c)}\Big) \Big], \,\, \mbox{ vasodilation} \label{eq:iplm40}
\end{align}
while microrotational velocity profiles for various segments are,
\begin{align}
    \displaystyle \omega_0 &= \frac{n_1 (u_0 - u_{slip})}{2 \, R(z)} \frac{\Big({\log(r)-\log(r_c)}\Big)}{\Big({\log(R(z)-\log(r_c)}\Big)^2}, \,\,\,\, \mbox{at normal segments}\label{eq:iplm41} 
    \\
  \omega_0 &= \frac{n_1}{2}\frac{\Big({\log(r_c) - \log(r)}\Big)}{\Big({\log(R(z)-\log(r_c)}\Big)} \,\,\,\,\, \mbox{at abnormal segments}. \label{eq:iplm421}
\end{align}
Likewise, by comparing the coefficients of higher powers of $p$ with the appropriate boundary conditions, we can obtain the solution. $p \rightarrow 1$, this process yields a semi-analytical solution for the variables of interest.
\begin{equation}
     S(r)  = \lim_{p \rightarrow 1} S_h = S_0 + S_1 + S_2 + S_3 + \cdots, \label{eq:iplm42}  
\end{equation}
where the variable $S\subset$ $\left( \Theta, \varsigma, v, \omega \right)$. 
\section{Results and Discussion}
\label{sec:4}
In this study, we present a mathematical model to gain insights into blood flow through a diseased model. Our analysis includes the calculation of velocity, temperature, and nanoparticle concentration profiles, along with studying the flow field characteristics using nondimensional parameters. The results are visually depicted through graphical representations.
\subsection{\textbf{Method validation and convergence of method}}
\label{4.1}
\begin{figure}[htbp]
\begin{center}
\includegraphics[width=0.9\textwidth,height=2.5in]{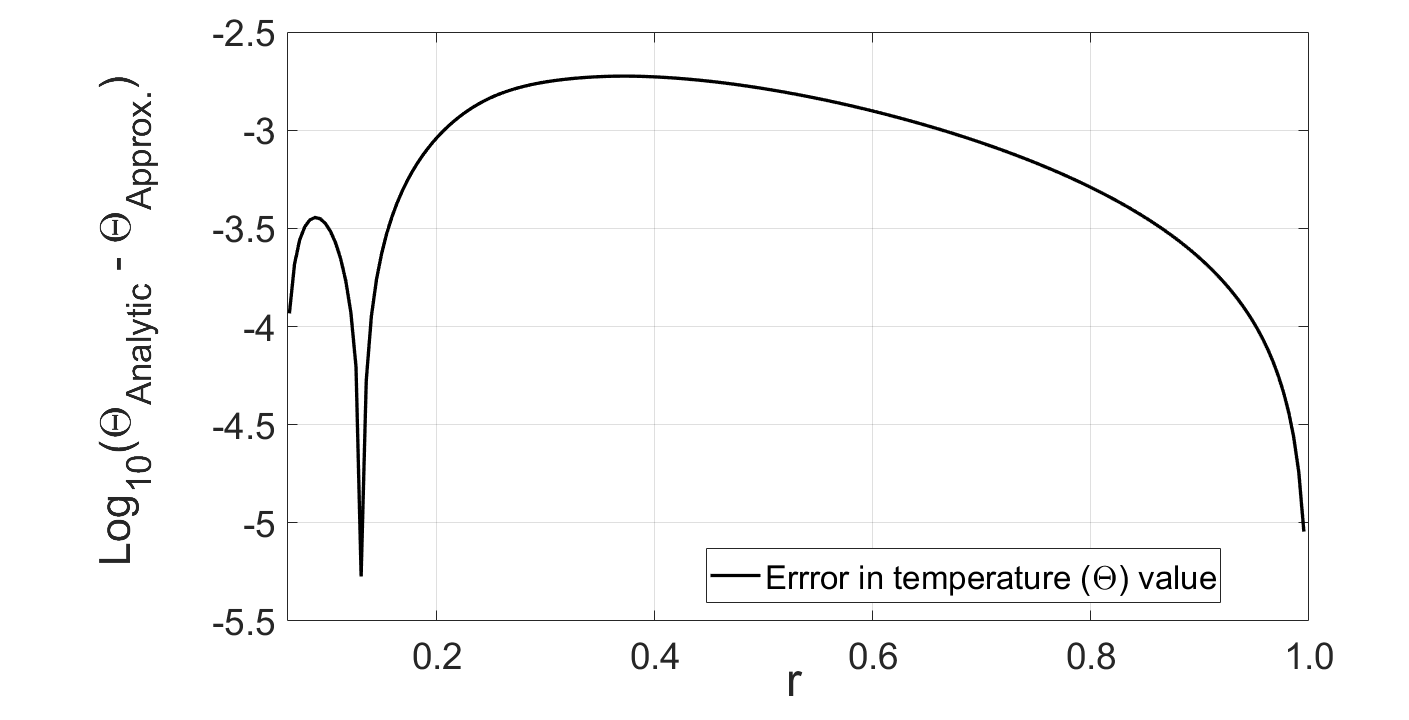}
\end{center}
\caption{Validation of HPM using temperature values.}
\label{fig2:validation}
\end{figure}
\begin{figure}[htbp]
\begin{center}
\includegraphics[width=0.8\textwidth,height=2.7in]{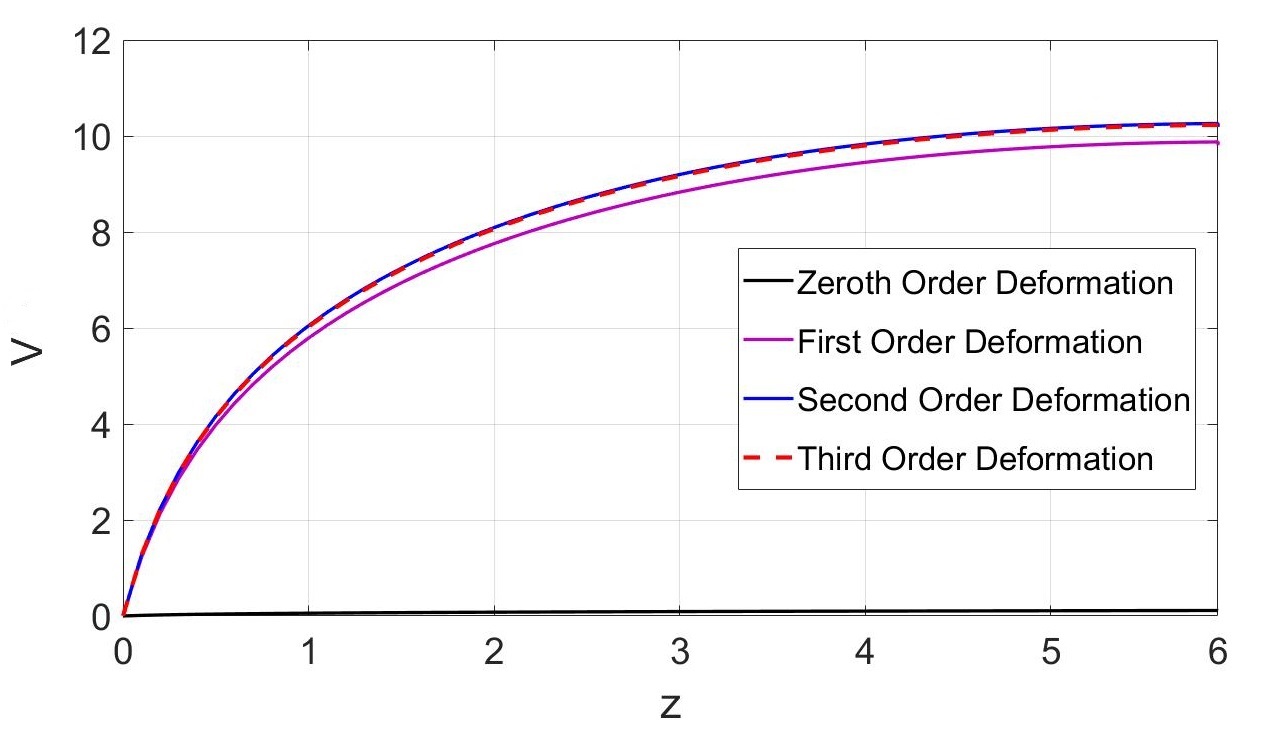}
\end{center}
\caption{Convergence of HPM using various orders of deformation of axial velocity.}
\label{fig2:convergence}
\end{figure}

Figures~\ref{fig2:validation} and \ref{fig2:convergence} depicts the validation and convergence of HPM, respectively. The logarithmic error values along radial direction were computed using the formula, $$\textit{Absolute Error} = \left(\Theta_{\mbox{Analytical}}-\Theta_{\mbox{Approx.}}, \right)$$ where the analytical temperature values are taken from the study~\cite{elnaqeeb2016cu1} and temperature values are computed using our considered HPM as shown in Fig.~\ref{fig2:validation}. The obtained absolute error values of $\mathcal{O}(10^{-3})$ validate the reliability of the HPM for solving non-linear coupled PDEs. It is worth noting that this perturbation method requires careful consideration of a series of deformations to achieve approximate solutions for the variables.
The convergence of the HPM method is depicted in Fig.~\ref{fig2:convergence}, where different orders of deformations of axial velocity were used.  We observed that there is a negligible difference between  the $2^{nd}$ and $3^{rd}$ orders of deformations in the axial velocity, while a significant difference exists between the $1^{st}$ and $2^{nd}$ orders of deformations, demonstrating the convergence of the method. Henceforth, used the $3^{rd}$ order of deformation to compute the solution of the variables in this study. 
\subsection{\textbf{Effect of $N_b$ and $N_t$ parameters}}
\label{4.2}
\begin{figure}[htbp]
\begin{center}
\includegraphics[width=0.495\textwidth,height=2.15in]{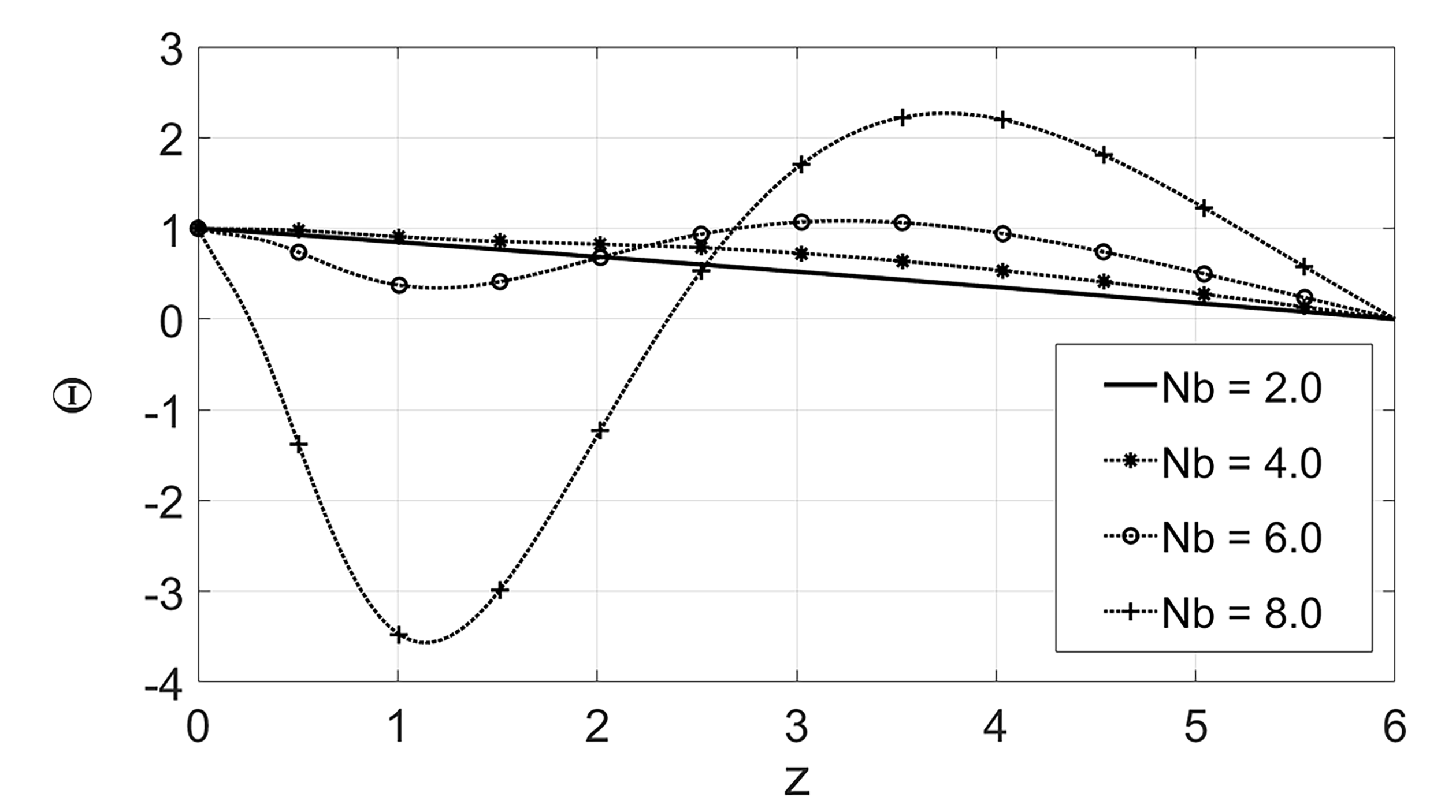} 
\includegraphics[width=0.495\textwidth,height=2.15in]{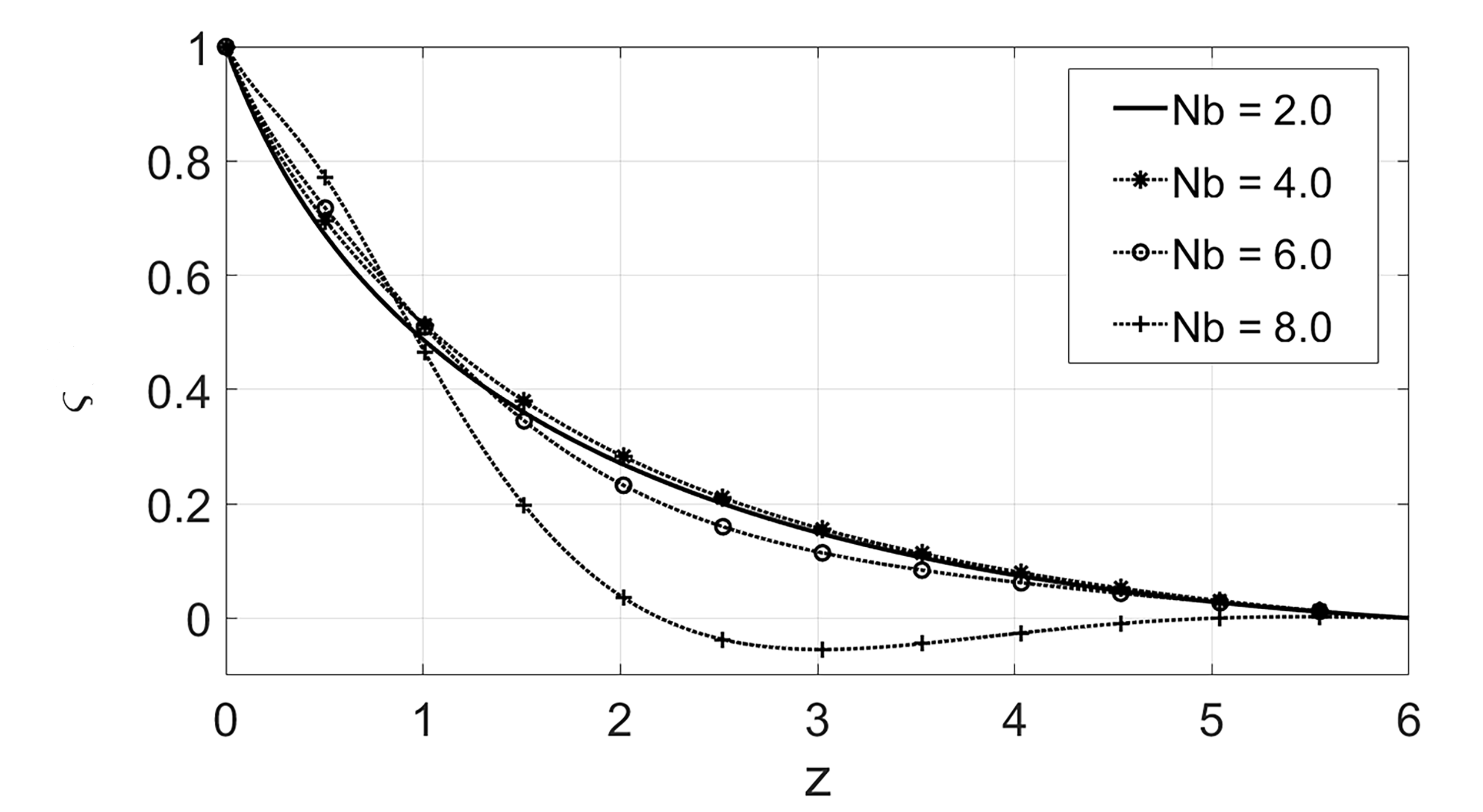}
\end{center}
\hspace{1cm}(a) Temperature Distribution \hspace{1.25cm} (b) Concentration Distribution
\caption{The influence of $N_b$ values on $\Theta$ and $\varsigma$ distribution in $z-$ direction.}
\label{fig3:thsg_nb}
\end{figure}

\begin{figure}[htbp]
\begin{center}
\includegraphics[width=0.495\textwidth,height=2.15in]{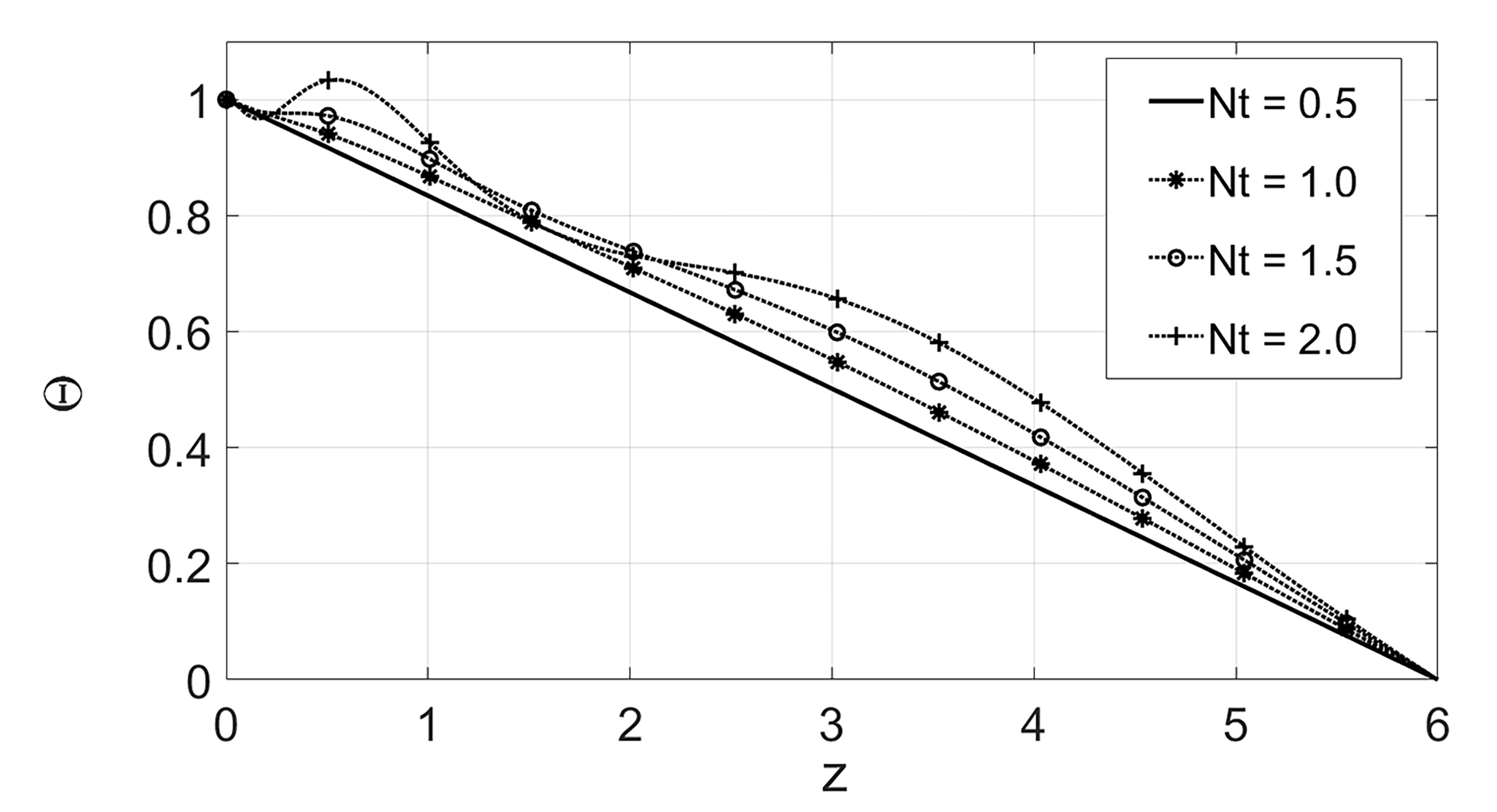} 
\includegraphics[width=0.495\textwidth,height=2.15in]{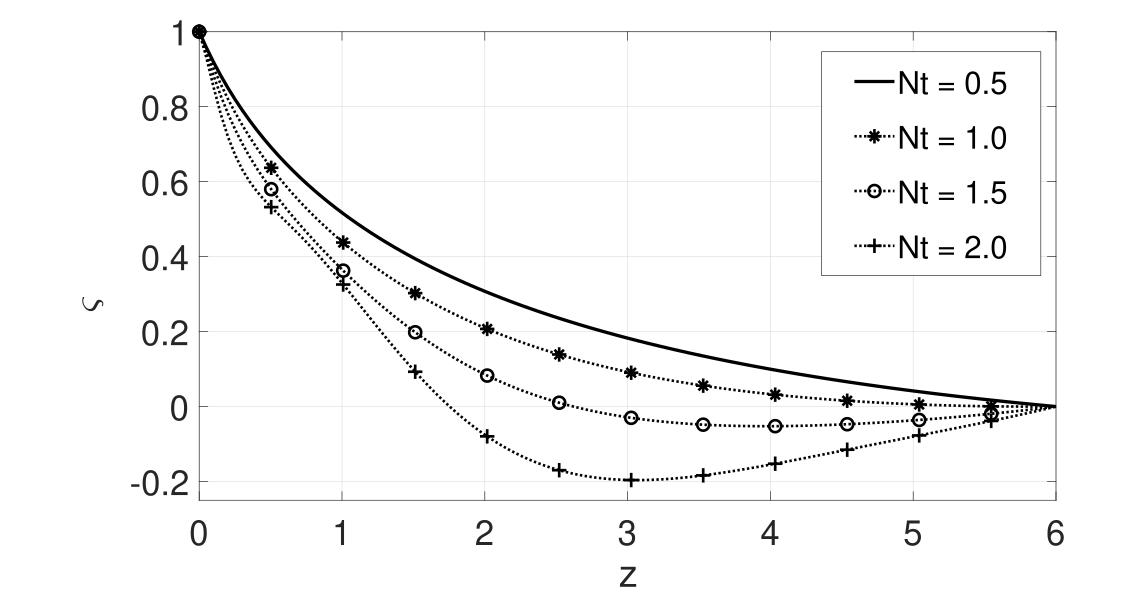}
\end{center}
\hspace{1cm}(a) Temperature Distribution \hspace{1.25cm} (b) Concentration Distribution
\caption{The influence of $N_t$ values on $\Theta$ and $\varsigma$ in $z-$ direction.}
\label{fig4:thsg_nt}
\end{figure}

The influence of non-dimensional parameters $N_{b}$ and  $N_{t}$ on non-dimensional temperature ($\Theta$) and concentration ($\varsigma$) values are shown in Figs.~\ref{fig3:thsg_nb} and \ref{fig4:thsg_nt}, respectively. Figure~\ref{fig3:thsg_nb} shows that an increase in $N_{b}$  leads to a corresponding increase in both the temperature and concentration profiles until $N_b = 4.0$. We observed that, in the vasoconstriction and vasodilation segments, respectively, the temperature decreases and increases rapidly whereas the concentration decreases gradually. This behavior can be attributed to the molecules experiencing bombardment from the surrounding fluid, causing an increase in the $N_{b}$ parameter. Consequently, the immersed particles exhibit significant, erratic movements, resulting in elevated temperature and concentration values. Therefore, the dispersion associated with the Brownian motion parameter $N_{b}$ significantly influences these phenomena.
Thermophoresis is a phenomenon that occurs when particles are randomly moving in a fluid, and the Brownian motion is the one where some variables experience small and random variations. As shown in Fig.~\ref{fig4:thsg_nt}, the increase in $N_t$ values is inversely related to the concentration and concentration dispersion, whereas it is directly proportional to temperature and temperature dispersion. Beyond a specific threshold value of $N_{t} = 1.5$, the temperature and concentration dispersion change unevenly due to the thermal conductivity and drug discharge effects. The temperature difference causes the particles to generate heat and disperse away from the surface of the catheter, along with the drug. It should be noted that the distribution of nanoparticles increases as the Brownian motion parameter value increases, whereas the trend is the opposite as the thermophoresis parameter value increases. As a result, for these two factors, the concentration profiles exhibit opposite behavior.

The Nusselt number $\left(Nu \right)$ is the ratio of conductive to convective heat transfers at the catheter wall. It is expressed in the dimensionless form as shown below: 
\begin{equation}
\displaystyle Nu = -\left( \frac{\partial \Theta}{\partial r} \right)_{r = r_{c}}.
\label{eq:p8}
\end{equation}
The Sherwood number, which is the ratio of diffusive mass transport to convective mass transfer in the context of the mass transfer phenomenon, is expressed below:
\begin{equation}
\displaystyle Sh = -\left( \frac{\partial \varsigma}{\partial r} \right)_{r = r_{c}}.
\label{eq:p9}
\end{equation}

\begin{figure}[htbp]
	\centering
	\includegraphics[width=0.9\textwidth,height=2.5in]{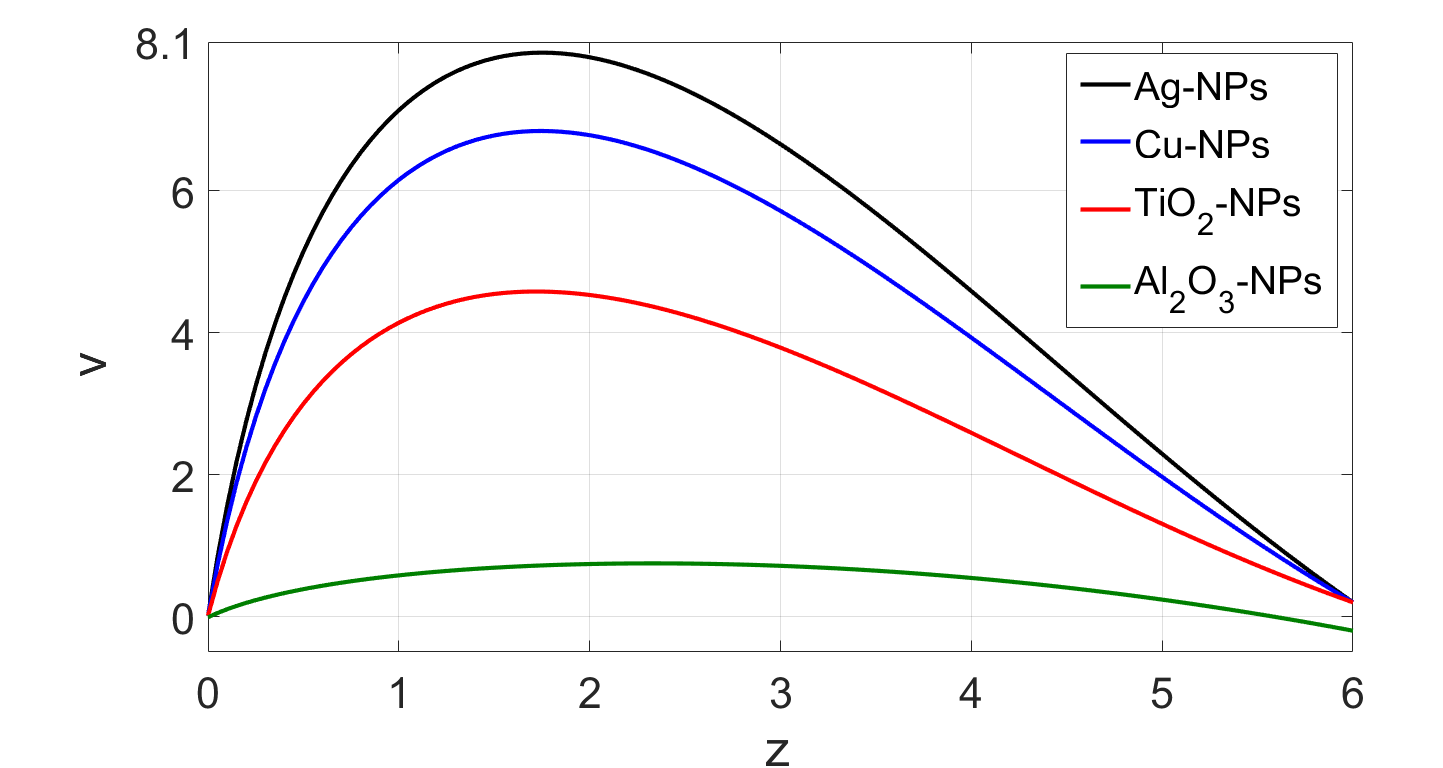}
	\caption{The influence of various NPs on axial velocity}
	\label{fig6:vel_nps}
\end{figure}

\begin{figure}[htbp]
	\centering
	\includegraphics[width=0.9\textwidth,height=2.5in]{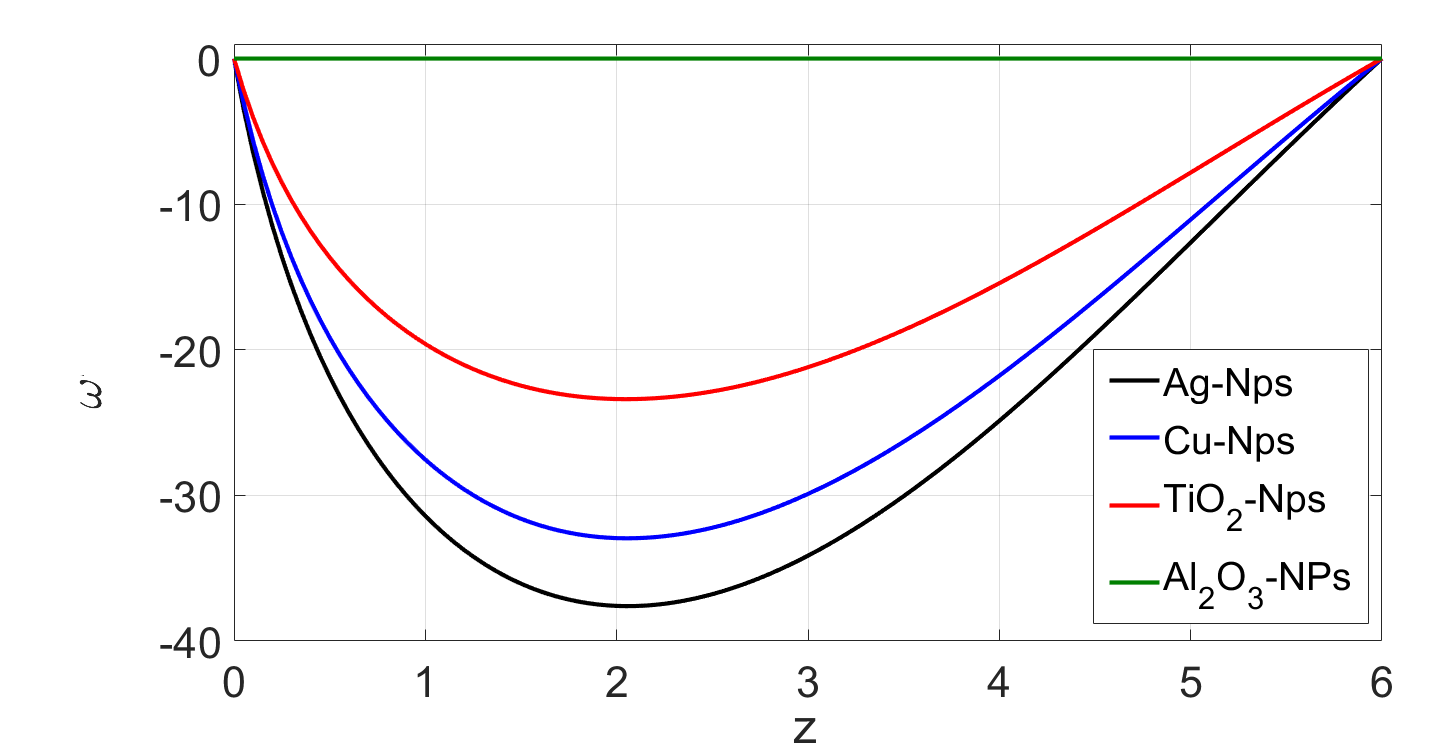}
	\caption{ The influence of various NPs on micro-rotational velocity}
	\label{fig7:omega_nps}
\end{figure}

\begin{figure}[htbp]
	\begin{center}
	\includegraphics[width=0.9\textwidth,height=2.5in]{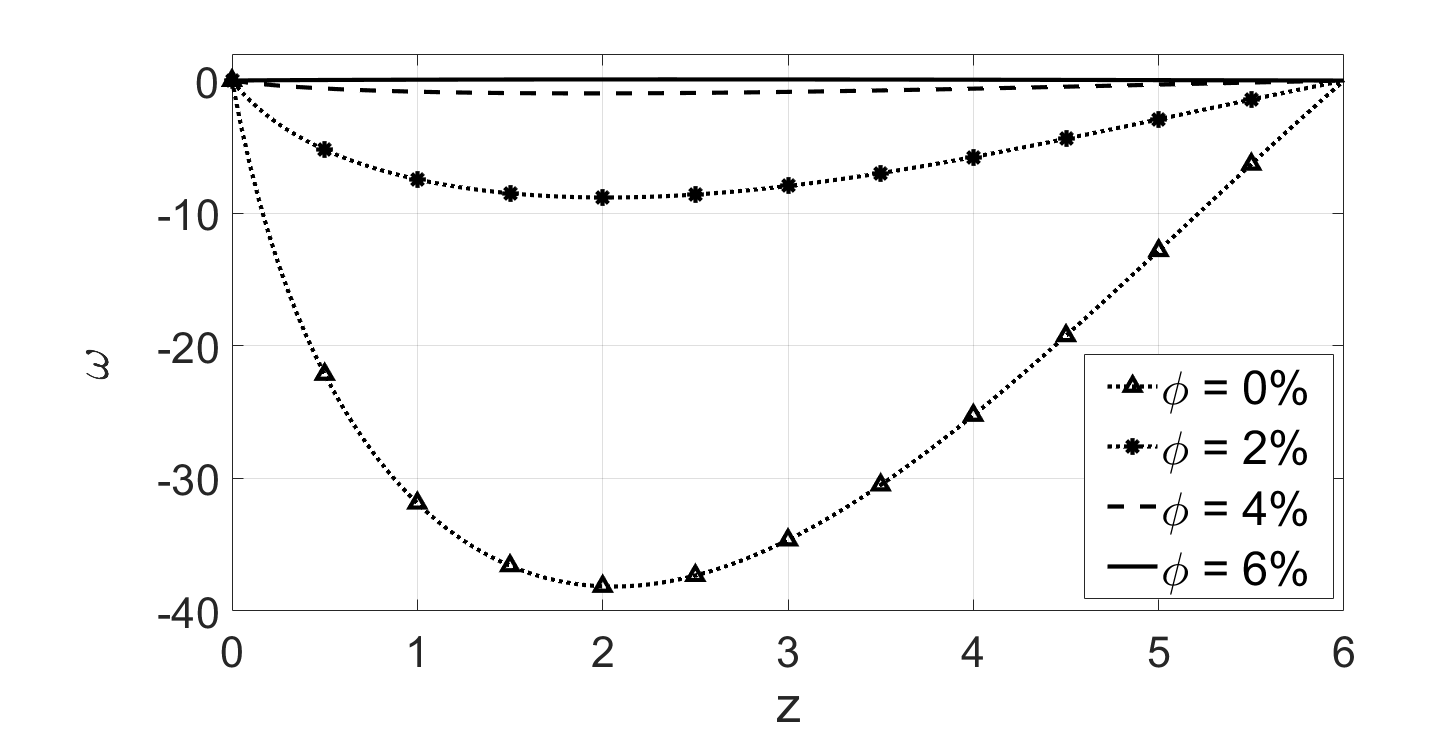}
	\end{center}
	\caption{The influence of various $\phi$ values on micro-rotational velocity.}
	\label{fig8:phi_omega}
\end{figure}

Tables~\ref{tab1:NuSh_Nb} and \ref{tab2:NuSh_Nt} provide the average Nusselt number and Sherwood number values corresponding to different values of $N_b$ and $N_t$, respectively. The average Nusselt number increases with an increase in $N_b$, implying that convective heat transfer becomes more dominant with higher Brownian motion parameter values, as shown in table \ref{tab1:NuSh_Nb}. On the other hand, the average Sherwood number also increases with an increase in $N_b$, indicating that convective mass transfer is more significant with higher Brownian motion parameter values. Regarding $N_t$, the average Nusselt number increases with an increase in $N_t$, which suggests that temperature-dependent heat transfer becomes more pronounced with higher thermophoresis parameter values, as mentioned in table \ref{tab2:NuSh_Nt}. In contrast, the average Sherwood number decreases with an increase in $N_t$, implying that concentration-dependent mass transfer becomes less dominant with higher thermophoresis parameter values. Finally, these observations highlight the important roles played by the Brownian motion parameter and thermophoresis parameter in influencing heat and mass transfer processes, respectively, within the diseased vascular model studied.
\begin{table}[htbp]%
\centering
\renewcommand{\baselinestretch}{1.5}\normalsize
\begin{tabular}{|c|c|c|c|c|} \hline
{} &{} & {$N_{t}$} & {${Nu}$} & {${Sh}$} \\ \hline
\multirow{4}{*}{$N_{b}$} & 1.0 & \multirow{4}{*}{0.7} & 1.2958 & 1.8149 \\  \cline{4-5}
 &  2.0 & & 1.3188 & 1.7215 \\ \cline{4-5}
 & 3.0 & & 1.4034 & 1.6827 \\ \cline{4-5}
& 4.0 & & 1.5461 & 1.6639\\  \cline{4-5}
& 5.0 & & 1.9478 & 1.6449 \\  \cline{4-5}
\hline
\end{tabular}
\caption{Average $Nu$ and $Sh$ for different $N_b$ values.}
\label{tab1:NuSh_Nb}
\end{table}
\begin{table}[htbp]
\centering
\renewcommand{\baselinestretch}{1.5}\normalsize
\begin{tabular}{|c|c|c|c|c|} \hline
{} &{} & {$N_{b}$} & {$Nu$} & {$Sh$} \\ \hline
\multirow{4}{*}{$N_{t}$} & 0.0 & \multirow{4}{*}{2.0} & 1.2939 & 1.5117 \\  \cline{4-5}
 &  0.5 & & 1.3081 & 1.6481 \\ \cline{4-5}
 & 1.0 & & 1.3394 &  1.8518\\ \cline{4-5}
& 1.5 & & 1.3972 & 2.1231  \\  \cline{4-5}
& 2.0 & & 1.6214 & 2.52815  \\  \cline{4-5}
\hline
\end{tabular}
\caption{ Average $Nu$ and $Sh$ for different $N_t$ values.}
\label{tab2:NuSh_Nt}
\end{table}
Figures~\ref{fig6:vel_nps} and \ref{fig7:omega_nps} illustrate the influence of different NPs on axial and micro-rotational velocities along the axial direction, respectively. In contrast to other NPs,$Ag-$NPs exhibit a high axial velocity compared to other NPs. However, due to the boundary conditions employed in the analysis, the micro-rotational velocity for $Ag-$NPs is relatively low. Accordingly, we considered $TiO_{2}-$NPs for further investigation, given their applications in photo-electro-chemical and biomedical fields, including antibacterial surface coatings and enhanced detection of targets, as highlighted in the studies ~\cite{jafari2020, ziental2020}.  Figure~\ref{fig8:phi_omega} further demonstrates the influence of NP volume fraction ($\phi$) on micro-rotational velocity.  We have observed that the micro-rotational velocity is negatively proportional to the velocity gradient in the radial direction. Therefore, we realize that the velocity increases as $\phi$ rises in value. To maintain the continuum of fluid, we have chosen $\phi=2\% $ even though we are aware that the micro-rotational velocity is substantially lower for higher values of $\phi$. This decision ensures that the fluid behavior remains consistent and aligns with real-physiological conditions.

\begin{figure}[htbp]
	\begin{center}
		\includegraphics[width=0.9\textwidth,height=2.5in]{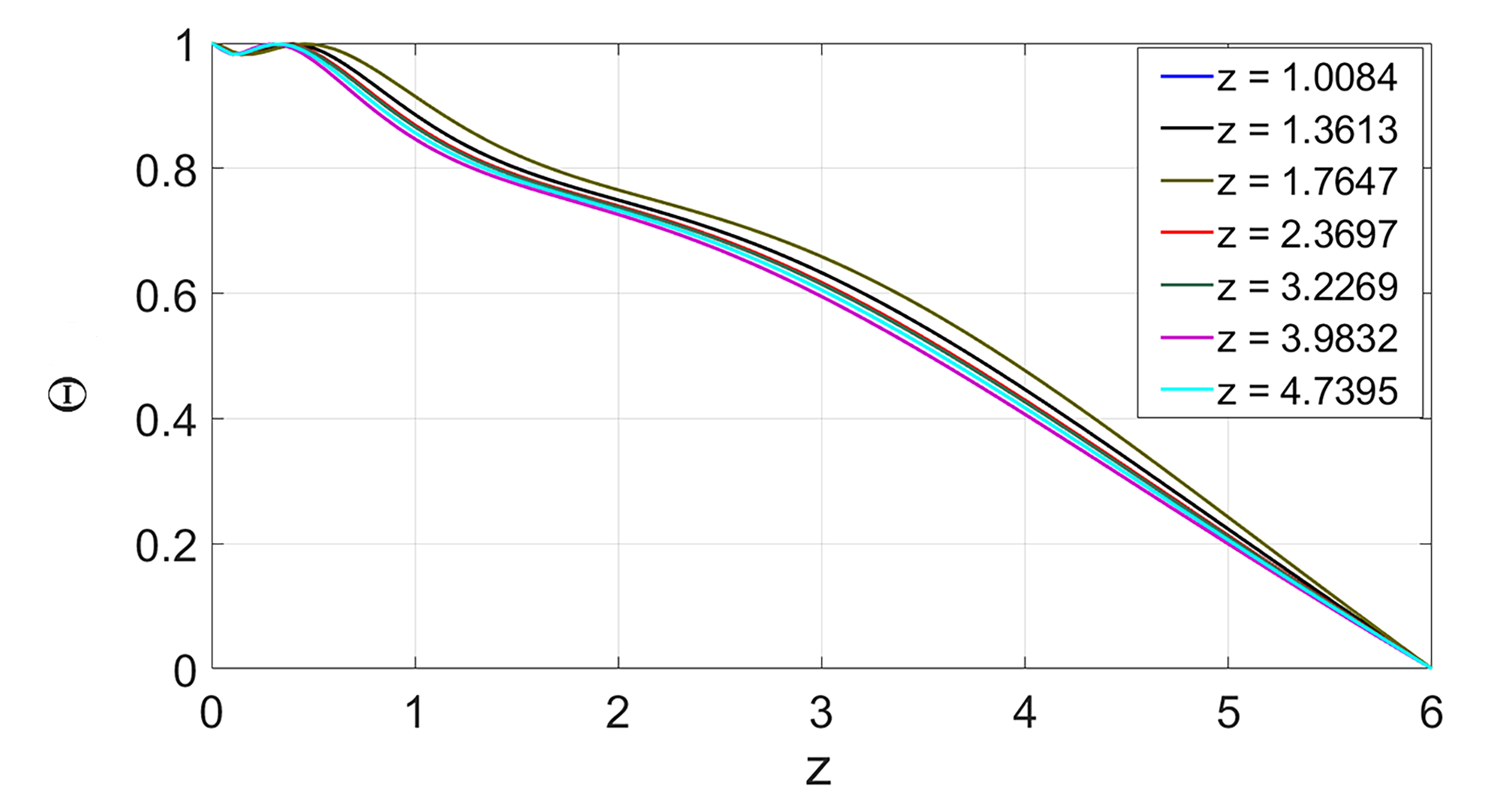} 
	\end{center}
	\caption{Distribution in $\Theta$ values with various $z-$ values.}
	\label{fig9:th_z}
\end{figure}

\begin{figure}[htbp]
	\begin{center}
	\includegraphics[width=0.95\textwidth,height=2.5in]{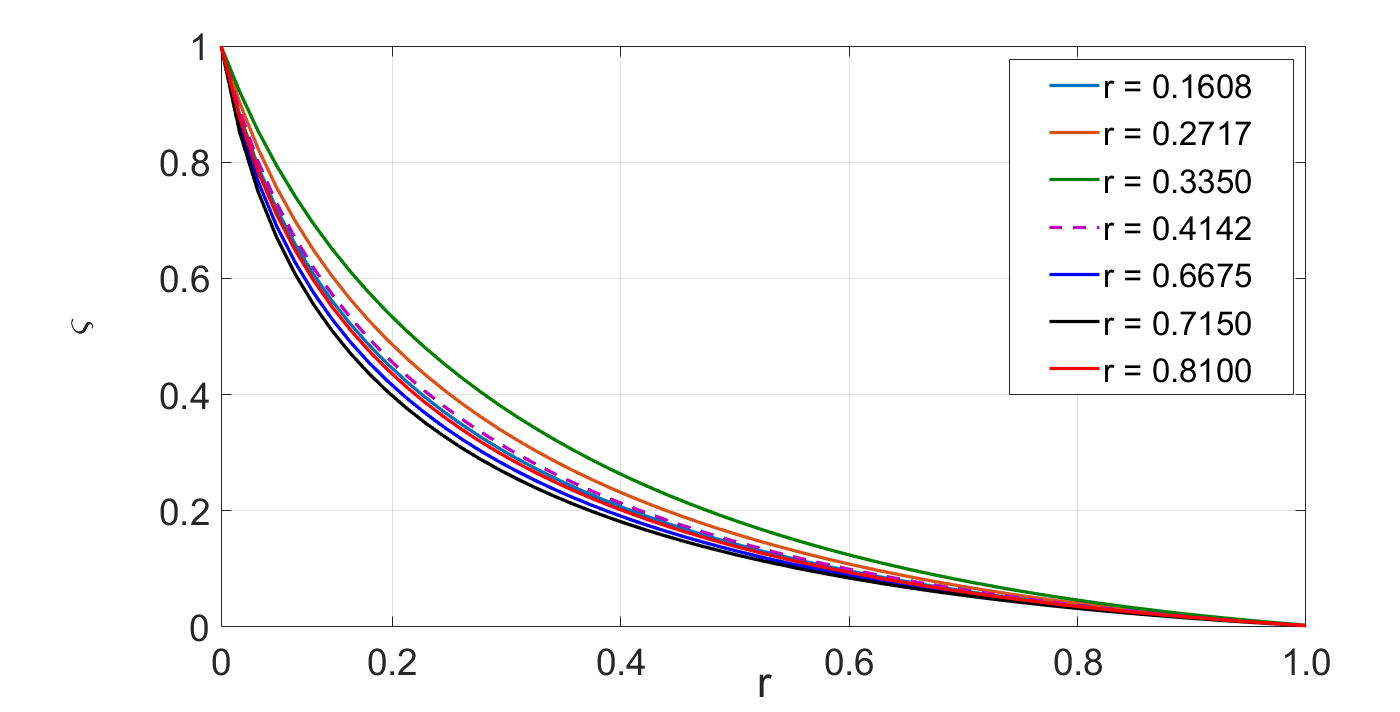} 
	\end{center}
	\caption{Distribution in $\varsigma$ values with various $r-$ values.}
	\label{fig10:sg_r}
\end{figure}

In Fig.~\ref{fig9:th_z}, the temperature values at various locations in the axial direction ($z$) are depicted. Notably, temperature fluctuations are particularly noticeable in the vicinity of the catheter wall and in the region where the blood vessels are constricted. Furthermore, we have observed that, compared to the regular segments, the temperature is lower in the vasoconstriction segment and higher in the vasodilation segment. These temperature variations are significant indicators of the effects of vascular conditions on blood flow and heat transfer. Figure \ref{fig10:sg_r} displays the variation in concentration for different annular regions. Due to the presence of NPs, fluid diffusion in the radial direction ($r$) from the catheter to the artery wall is efficient and homogeneous. The therapeutic NPs coated with drugs exhibit higher thermal conductivity as $\phi$ values increase. However, at higher volume concentrations, the anticipated improvement is reduced, accompanied by an increase in viscosity. These findings suggest that NP-drugs could be effective in detecting dysfunctional endothelium and can be beneficial for the prevention of CVDs. 
\subsection{\textbf{Flow field characteristics}}
\label{sec4.3}
\begin{figure}[htbp]
\begin{center}
\includegraphics[width=0.85\textwidth,height=2.5in]{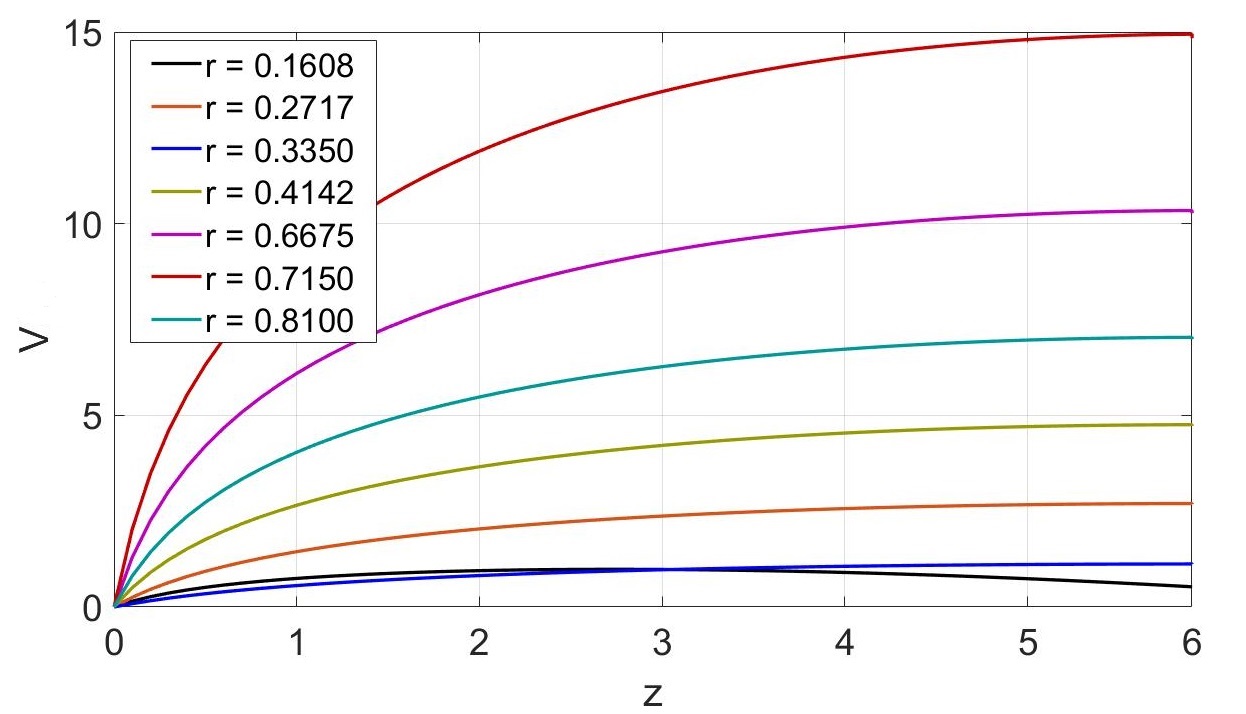} 
\end{center}
\caption{Variations in axial velocity for various $r-$ values. }
\label{fig11:vel_r}
\end{figure}

\begin{figure}[]
\begin{center}
\includegraphics[width=0.85\textwidth,height=2.5in]{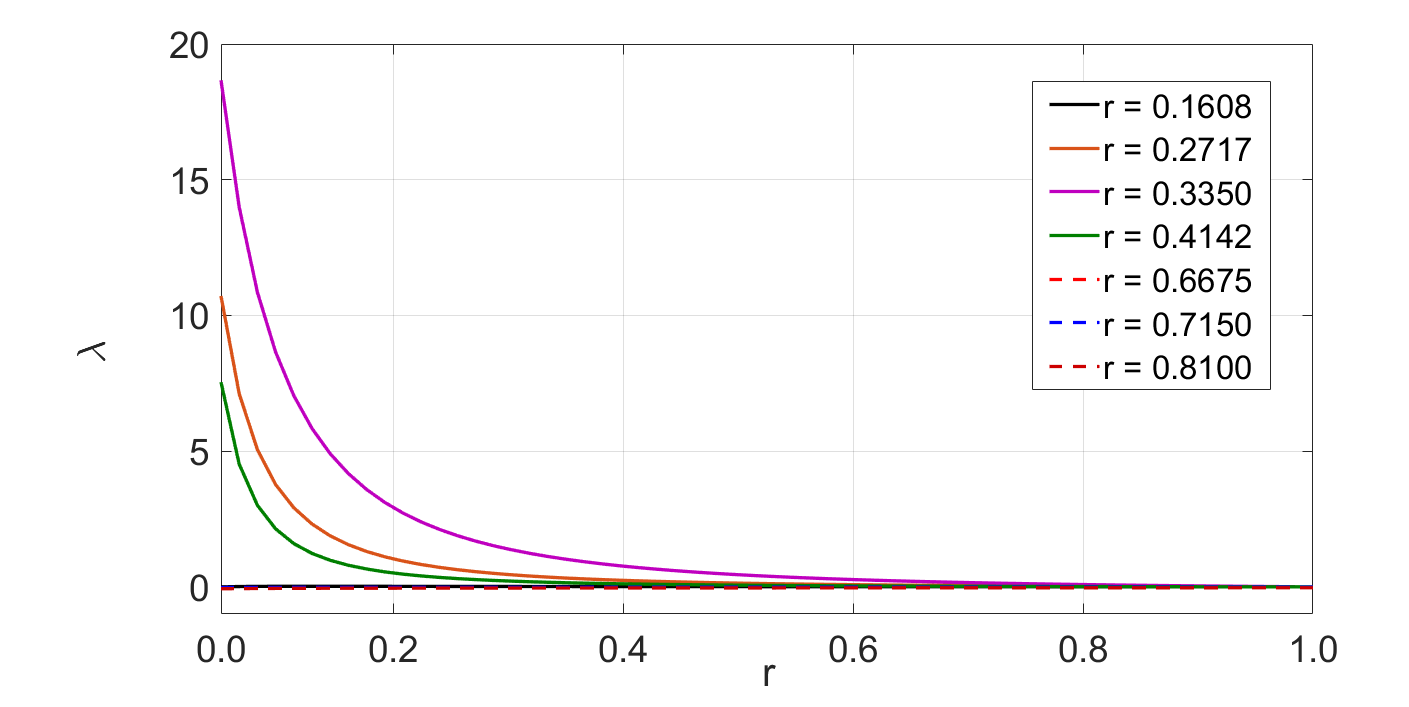}
\end{center}
\caption{Variation in impedance profiles for various $r-$ values.}
\label{fig19:Imp_r}
\end{figure}

In Fig.~\ref{fig11:vel_r}, the axial velocity profiles are displayed for varying annular radii. The axial velocity in the annular region is significantly low at the radius corresponding to the minimum constriction $(r = 0.3350)$ and closer to the catheter wall, while it is higher at the radius corresponding to the maximum dilation $(r = 0.7150)$. This observation indicates that the flow rate $\displaystyle \mbox{Q} = \int_{r_c}^{R(z)} (2,r,v,dr)$ increases from the constricted region to the vasodilation segment. The increase in flow rate is attributed to the increasing velocity gradient as the velocity profile rises. This behavior is consistent with the flow behavior and the varying vessel sizes encountered in the diseased  model.
The variations in impedance profiles due to varying annular radii were also understood from Fig.~\ref{fig19:Imp_r}.
The impedance ($\lambda$), a crucial fluid flow characteristic, is given as $\displaystyle \lambda = \frac{\Delta p}{Q},$  which is associated with the drop in pressure and rate of flow. This equation uses Eq. (\ref{ch2_eq28}) to compute the pressure drop across the diseased artery as follows:
\begin{multline}
\Delta P  = \int_{0}^{L} \Bigg[\left( C_{\mu}+\frac{N}{(1-N)}C_{\eta} \right)\,\left( {\partial_{rr} v} + \frac{1}{r} {\partial_r v} \right) + 2\left(\frac{N}{(1-N)} \right) \,C_{\eta}\,\left( {\partial_r \omega} + \frac{\omega}{r} \right) \\ + \left( \beta_{nf} G{r}\Theta + \beta_{nf} B{r}\,\varsigma\, \right) \Bigg] dz. 
\label{eq:p1}    
\end{multline}
The Impedance is high at the  constriction segment corresponding to $r = 0.1608$, and it decreases progressively from the constricted segment to the dilation segment, as illustrated in Fig.~\ref{fig19:Imp_r}. However, this dispersion is quite low for the segment. This phenomenon is inversely related to the axial velocity profiles, as illustrated in Fig.~\ref{fig11:vel_r}, implying that the rate of flow is what determines the flow field characteristics, rather than pressure drop. The fact that we are focusing on the transfer of drugs to treat the organ makes this situation quite beneficial.
\subsection{\textbf{Effect of $N$ and $M$ parameters}}
\label{sec4.5}
\begin{figure}[]
\centering
\begin{center}
\includegraphics[width=0.9\textwidth,height=2.5in]{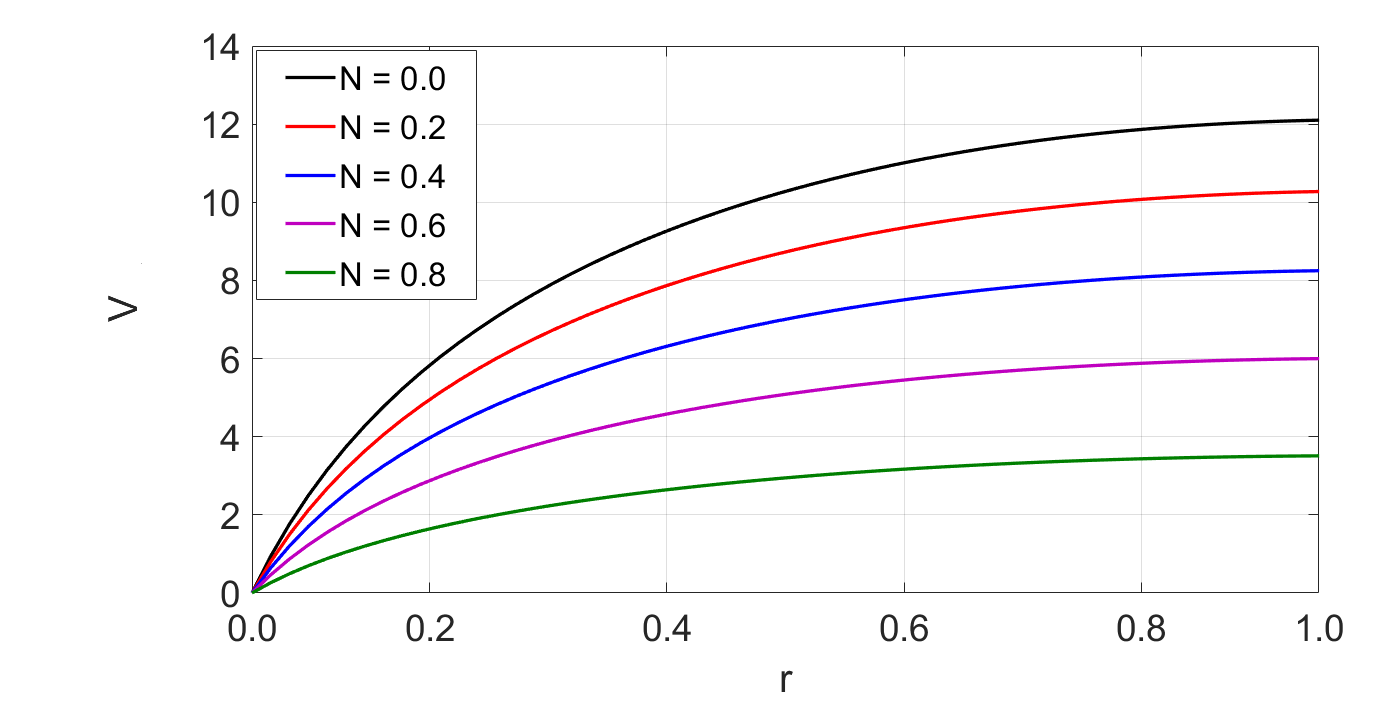}
\end{center}
(a) Axial Velocity
\begin{center}
\includegraphics[width=0.88\textwidth,height=2.5in]{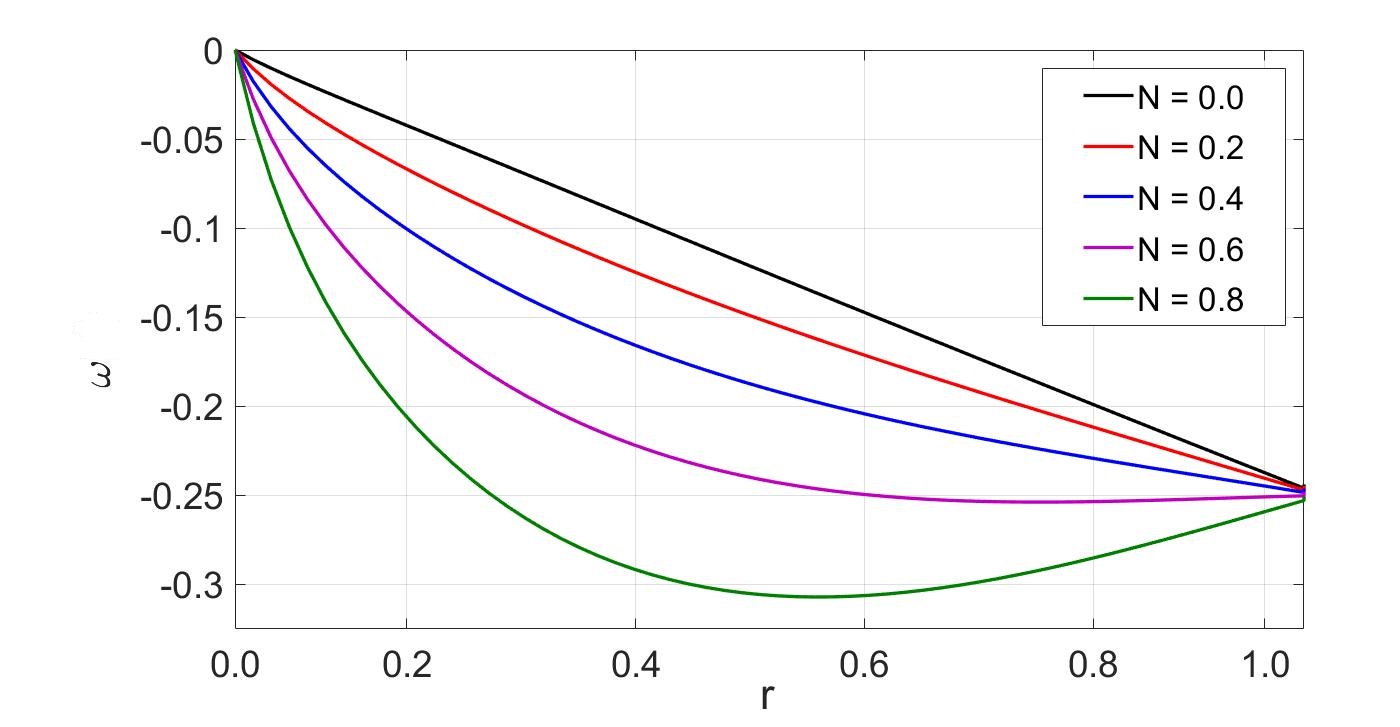}\\
(b) Micro-rotational Velocity
\end{center}
\caption{Effects of $N$ values on (a) axial  and (b) micro-rotational velocity profiles.}
\label{fig15:velomega_N}
\end{figure}
\begin{figure}[]
\begin{center}
\includegraphics[width=0.9\textwidth,height=2.75in]{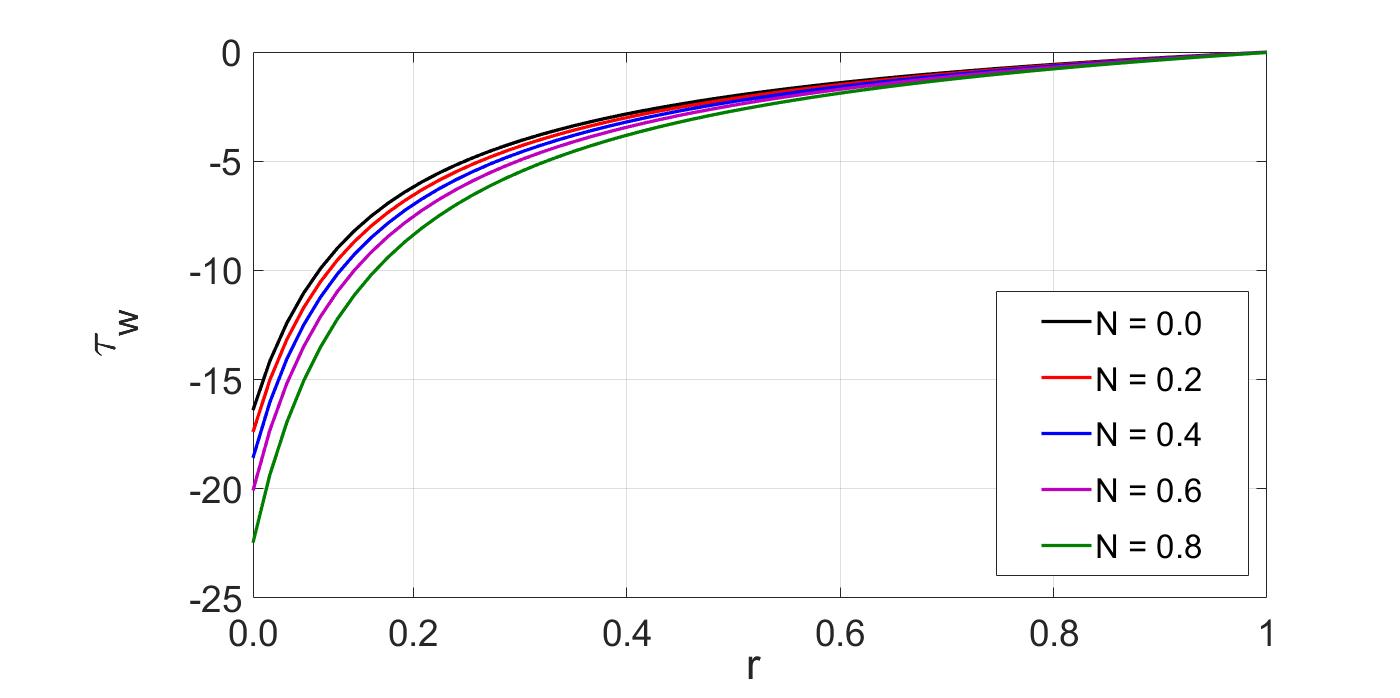}
\end{center}
\caption{Effects of $N$ values on WSS profiles.}
\label{fig18:WSS_N}
\end{figure}
\begin{figure}[]
\centering
\begin{center}
\includegraphics[width=0.85\textwidth,height=2.25in]{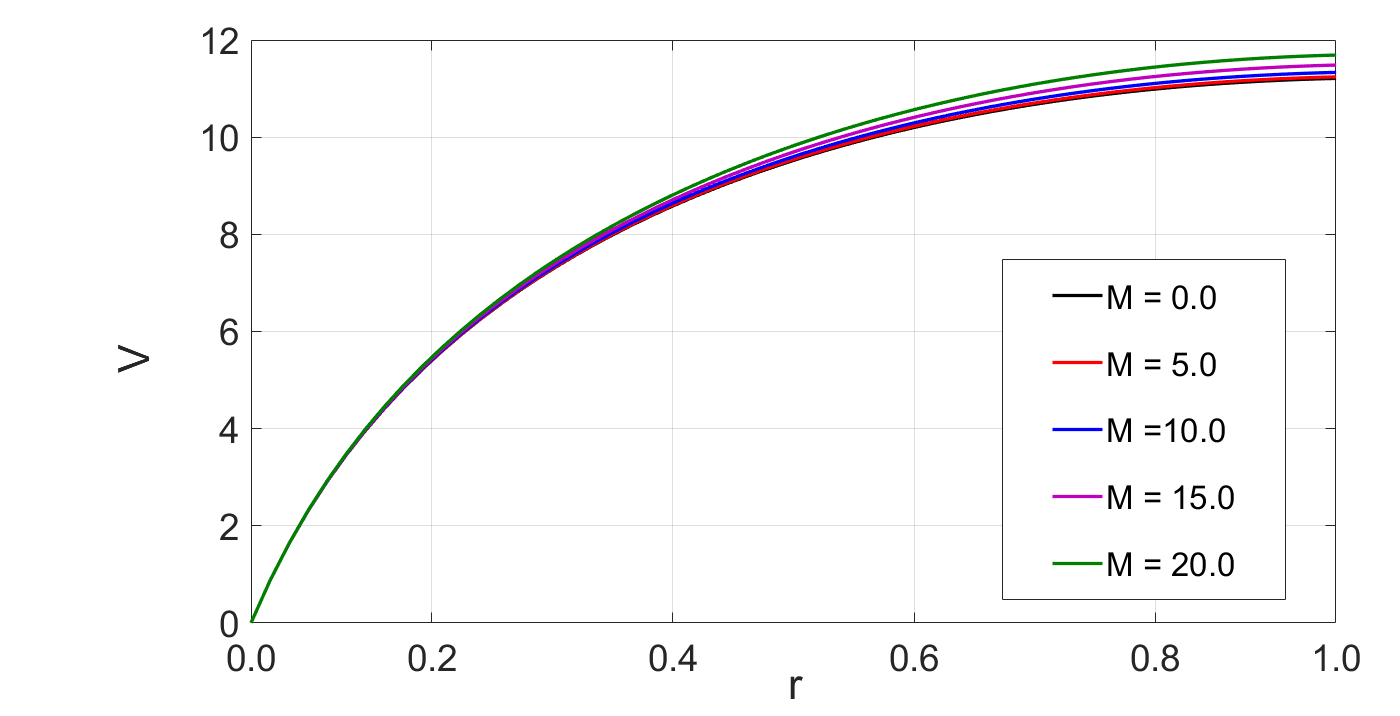}
\end{center}
(a) Axial Velocity
\begin{center}
\includegraphics[width=0.85\textwidth,height=2.4in]{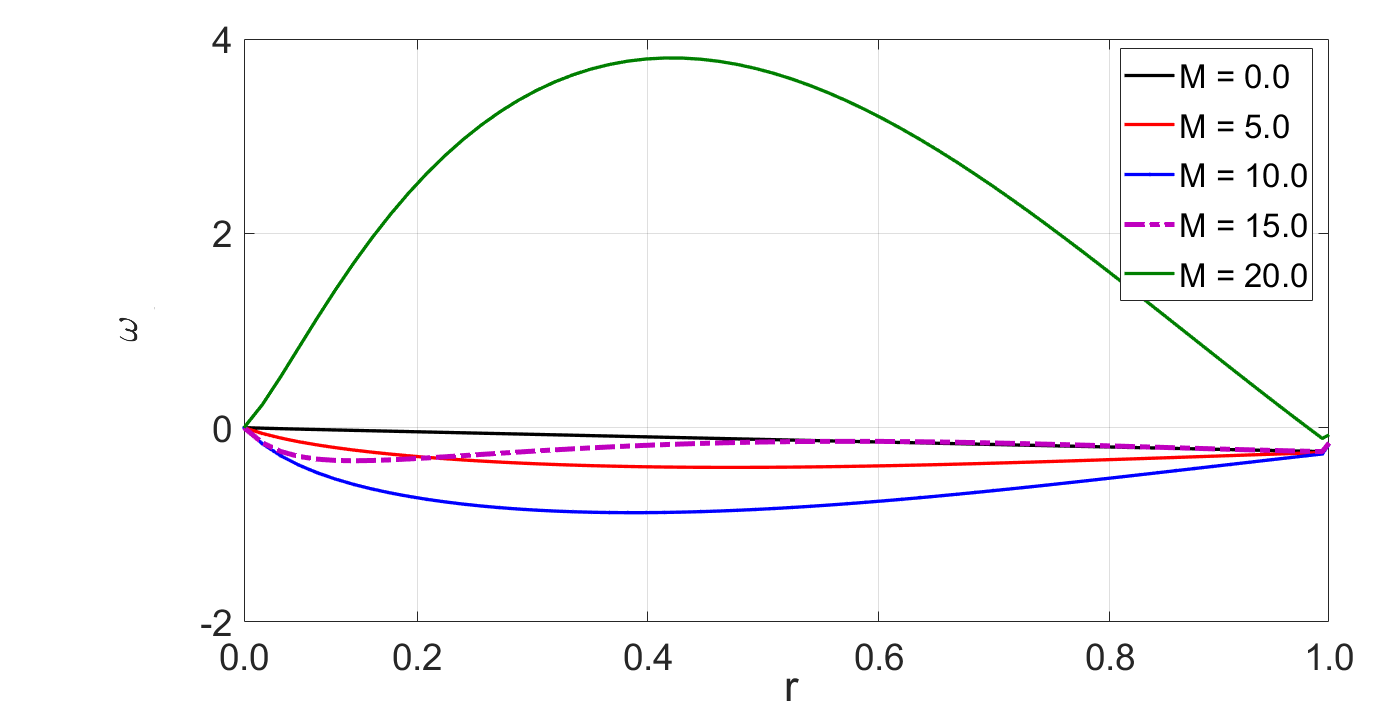}
\end{center}
(b) Micro-rotational Velocity
\caption{Effects of $M$ values on (a) axial  and (b) micro-rotational velocity profiles.}
\label{fig16:velomega_M}
\end{figure}
The effect of the coupling number $(N)$ and micropolar parameter $(M)$ on hemodynamic characteristics is depicted in Figs.~(\ref{fig15:velomega_N} -- \ref{fig16:velomega_M}). These parameters depend on the dynamic viscosity and vortex viscosity of the micropolar fluid. In the case of a micropolar fluid, the coupling number $N$ ranges in the interval $(0,1)$, with $N = 0$ representing a Newtonian fluid. As the coupling number $N$ increases, the fluid becomes more viscous, leading to increased friction between fluid layers, as shown in Fig.~\ref{fig15:velomega_N}. However, the magnitude of $\omega$ increases with the increasing values of $N$, which indicates the occurrence of secondary flow shown in Fig.~\ref{fig15:velomega_N}(b), as the fluid becomes more viscous.  

We have also understood the effect of $N$ on WSS from Fig.~\ref{fig18:WSS_N}. WSS is one of most the important characteristics to determine the resultant force between the blood flow and vessel wall, which is mathematically written for the micropolar fluid as;
\begin{equation}
\tau_{rz} = \frac{1}{(1-N)}\,\left( {\partial_r v}\right)_{r = R(z)}.
\end{equation}\label{eq:p5}
Based on the observations, we have found that WSS tends to be low for higher $N$ values and relatively high for smaller $N$ values. Selecting an appropriate value for the coupling number ($N$) is crucial, especially considering that blood is a non-Newtonian fluid. For this study, we opted to use $N=0.2$ for the computations, which best captures the non-Newtonian behavior of blood. The micropolar parameter ($M$) also plays a significant role in determining hemodynamic characteristics. An increase in $M$ leads to more pronounced micropolar fluid effects, resulting in changes in flow patterns and shear stresses. In Fig.~\ref{fig16:velomega_M}, we observed that the axial velocity profiles gradually rise as $M$ values increase, while the micro-rotational velocity exhibits erratic behavior until $M=15.0$, after which it rises rapidly for $M = 20.0$. This suggests that a lower value of $M$ generates the rotational flow, leading us to choose $M=15.0$ for the computations. With these observations, we have considered the appropriate values for the non-dimensional parameters $N$ and $M$, which are crucial for accurately representing the complex hemodynamic characteristics of the fluid flow in the vascular model. These parameter choices ensure that the simulation results align closely with real-world fluid behavior, which is vital for gaining insights into blood flow dynamics and developing effective drug delivery strategies.
\section{Conclusions}\label{sec:5}
The interest in understanding the thermophysical properties of nanofluids has been growing alongside recent advancements in theoretical modeling. Notably, it has been established that the migration of NPs has a significant impact on the movement and heat transfer behavior of nanofluids. By covering the catheter's external surface with NPs, we gain valuable insights into how blood nanofluid is transported through an arterial domain experiencing both vasoconstriction and vasodilation. Through this study, we analyzed the flow field characteristics and determined appropriate estimates for the non-dimensional parameter values, such as $\phi = 2\%$, $N = 0.2$, $M = 15.0$, $N_b = 4.0$, and $N_t = 1.5$. We observed an inverse relationship between the axial velocity profile and the impedance profile. The buoyancy force played a significant role in inducing natural convection within the nanofluid, resulting from temperature and concentration gradients in the blood flow. Moreover, NPs tend to aggregate when they are in contact, transported, or dispersed in blood. This research is vital for enhancing our understanding of NPs-driven transport phenomena in drug delivery, benefiting both the pathology and pharmaceutical industries. It sheds light on the intricate behavior of nanofluids and their potential applications in biomedical settings.
\begin{acknowledgements}
This work has been completed under the Institutional Fellowship of DIAT(DU), with Reference number: DIAT/F/Acad(PhD)/1613/15-52-13.
\end{acknowledgements}
\section*{Conflict of Interest} The co-authors have no potential conflicts of interest.
\bibliographystyle{spphys}       

\end{document}